\documentclass[preprint, amsmath, amssymb, showkeys, titlepage, superscriptaddress]{revtex4-2}
\usepackage{booktabs}
\usepackage{graphicx}
\usepackage{xcolor}
\usepackage{multirow}
\usepackage[bookmarks=true, colorlinks=true, citecolor=blue, linkcolor=blue, urlcolor=blue]{hyperref}
\usepackage{floatrow}
\usepackage{sidecap}
\usepackage{subcaption}
\usepackage{romannum}
\usepackage{rotating} 
\usepackage{ragged2e}
\usepackage{xparse}
\usepackage{tabularx}
\graphicspath{{figs/}}
\usepackage{graphicx}
\DeclareGraphicsExtensions{.pdf,.png,.jpg}

\begin{document}

\title{Diabetes Lifestyle Medicine Treatment Assistance Using Reinforcement Learning}

\author{Yuhan Tang}
\email{yuhan.tang@my.cityu.edu.hk}
\affiliation{Department of Data Science, College of Computing, City University of Hong Kong, Hong Kong, China}

\begin{abstract}
Type-2 diabetes prevention and treatment can benefit from personalized lifestyle prescriptions. However, the delivery of personalized lifestyle medicine prescriptions is limited by the shortage of trained professionals and the variability in physicians’ expertise. We propose an offline contextual bandit approach that learns individualized lifestyle prescriptions from the aggregated NHANES profiles of 119,555 participants by minimizing the Magni glucose risk–reward function. The model encodes patient status and generates lifestyle medicine prescriptions, which are trained using a mixed-action Soft Actor-Critic algorithm. The task is treated as a single-step contextual bandit. The model is validated against lifestyle medicine prescriptions issued by three certified physicians from Xiangya Hospital. These results demonstrate that offline mixed-action SAC can generate risk-aware lifestyle medicine prescriptions from cross-sectional NHANES data, warranting prospective clinical validation. 
\end{abstract}

\maketitle

\section{Related Work}
Lifestyle medicine focuses on preventing and reversing chronic conditions, such as type 2 diabetes and cardiovascular diseases, through improving critical lifestyle factors. These factors typically include diet, sleep, physical activity, stress management, cessation of smoking and alcohol consumption, and social interactions. Unlike traditional medical disciplines that primarily rely on pharmaceutical treatments, lifestyle medicine emphasizes personalized prescriptions based on a patient’s specific condition and lab test results related to nutrition, exercise, sleep, and psychological status. Such individualized care necessitates a high degree of professional expertise and frequent, timely feedback from healthcare providers.

Despite its potential effectiveness, lifestyle medicine currently represents only a minor portion of mainstream medical treatment programs. Correspondingly, the number of physicians with specialized training and qualifications in lifestyle medicine remains limited \cite{Almulhim2024}. This gap contributes to insufficient medical resources and inconsistency in treatment outcomes. To address these challenges, our proposed model enables healthcare providers with varying levels of expertise to make scientifically grounded, consistent decisions when managing complex patient cases. Ultimately, this approach aims to bridge knowledge gaps across medical specialties and significantly improve the overall quality and effectiveness of patient care.

Artificial Intelligence has become a powerful tool in medical assistance, significantly enhancing the diagnosis and treatment of various diseases. Supervised learning models have been widely utilized for disease diagnosis \cite{Dive2024,Nandan2024,Qin2022}. However, these methods heavily depend on the availability of large, accurately labeled datasets, which can be challenging to obtain. As an alternative, reinforcement learning, as a type of machine learning, involves the agent interacting with the environment and making decisions based on the rewards provided by the environment to achieve optimal outcomes \cite{Komorowski2018,Sun2021,Wang2023,Zheng2021}. This approach is particularly advantageous for diabetes prevention and management, as patients exhibit diverse physiological characteristics, such as age, BMI, and blood pressure, as well as varied lifestyle habits, all of which significantly influence blood glucose fluctuations \cite{Yau2023,Kannenberg2024}. Therefore, the ability to dynamically receive feedback on blood glucose changes from the environment would be highly important.

Diabetes is classified into two types: type 1 and type 2. Unlike type 2 diabetes, which typically results from acquired lifestyle factors, type 1 diabetes is caused by congenital immune disorders that prevent insulin production, thus requiring pharmacological interventions for insulin control. Current research on type 1 diabetes mostly focuses on closed-loop glucose control systems. For example, Magni employed Model Predictive Control (MPC) to accurately predict blood glucose fluctuations by simulating meal intake and metabolic dynamics, enabling the artificial pancreas to reduce the time error of insulin initiation \cite{Magni2007}. Javad developed a reinforcement learning (RL) model utilizing patient data including glycated hemoglobin (HbA1c), BMI, exercise, and alcohol consumption to dynamically adjust insulin dosage according to blood glucose responses \cite{OroojeniMohammadJavad2019}. In comparison to Javad’s model, our approach includes additional patient demographic information to enhance prediction accuracy. Furthermore, while Javad’s method uses the difference between actual and target glucose levels as the reward function, we extend this by incorporating risk assessments associated with varying glucose levels. Thus, our model not only seeks to achieve target blood glucose levels but also minimizes patient risk from hyperglycemia and hypoglycemia. 

Similarly, Fox \cite{Fox2020} and Viroonluecha \cite{Viroonluecha2022} used wearable glucose monitor data \cite{Zhu2021} to establish an RL model based on the Soft Actor-Critic (SAC) algorithm, adjusting insulin dosages according to real-time blood glucose readings and meal information. Like our approach, Fox adopted the Magni risk function as the reward criterion. However, in contrast to our contextual bandit \cite{Lei2017, Oh2022} model, which optimizes blood glucose management within each iteration without relying on subsequent states, Fox’s model used the original MDP. This is because, for type 1 diabetes, strictly controlling real-time blood glucose levels is the top priority. While lifestyle medicine prescriptions require patients to follow them for a period of time to achieve blood glucose control, underscoring the importance of maximizing the effectiveness of each individual prescription. Emerson's work demonstrated that offline RL models require fewer training samples and longer target blood glucose durations compared to online RL models, and are less tend to extreme action outputs \cite{Emerson2023}. However, the limitations of offline RL models are also evident, as their training sets are fixed and cannot interact with the environment.

Type 2 diabetes, compared to type 1, is particularly suitable for prevention and management through lifestyle medicine interventions. Previous studies have shown that changing the lifestyle of high-risk groups, such as diet and exercise, can effectively prevent type 2 diabetes \cite{Jaakko2025}. However, most existing studies \cite{Yom-Tov2017,Lauffenburger2021,Umar2024,Hohberg2022} using reinforcement learning algorithms have only proven that personalized guidance will greatly improve patient compliance and intervention effectiveness, but only as an incentive. Di's work \cite{Di2022} further applies this to the field of lifestyle medicine, producing lifestyle medicine prescriptions that are more theoretically sound. Di's model considers the degree of blood sugar reduction and quality of life (QoL) indicators as joint rewards. However, a common problem in current research \cite{Jafar2024}, including Di's, is the small sample size, and our separate training and validation sets of thousands of records make the results more convincing.

\subsection{Data Source}
\subsubsection{Training and testing dataset}
NHANES Survey dataset: National Health and Nutrition Examination Survey. It's conducted by the Centers for Disease Control and Prevention to measure the health and nutrition of Americans of all ages. The NHANES dataset combines information on demographics (age, sex, race/ethnicity, income, education), self‑reported behaviors (dietary intake, smoking, exercise), clinical measurements (body measurements, dental exams), and laboratory results (nutrient levels, cholesterol, glucose, environmental chemical exposures). The dataset covers the data from 1999 to 2023.

\subsubsection{Intervention dataset}
Data is from Xiangya Hospital’s Wellness Center and includes both physiological indicator data (e.g., blood test results, blood pressure, blood lipid levels, height, weight, etc.) and intervention plan data (e.g., exercise prescriptions, nutritional prescriptions, etc.). For example, if a patient shows signs of diabetes during a consultation, they will be enrolled in the center’s diabetes prevention program. In this program, physicians issue prescriptions addressing multiple aspects such as exercise, diet, sleep, and psychological health. The patient then adjusts their lifestyle according to these prescriptions and returns for a follow-up examination after several months. At that time, the physician will reassess the test results and the implementation of the previous prescriptions, making necessary adjustments to help the patient effectively control diabetes.

\section{Data}
\begin{table}[h]
    \centering
    \begin{tabular}{lccc}
        \toprule
        \textbf{Year} & \textbf{Total Sample Size} & \textbf{Diabetes} & \textbf{Prediabetes} \\
        \midrule
        1999-2000&9,965&489&58  \\
        2001-2002&11,039&528&78  \\
        2003-2004&10,122&559&78  \\
        2005-2006&10,348&521&189  \\
        2007-2008&10,149&777&242  \\
        2009-2010&10,537&739&262  \\
        2011-2012&9,756&708&245  \\
        2013-2014&10,175&737&278  \\
        2015-2016&9,971&856&513  \\
        2017-2020&15,560&1,445&952  \\
        2021-2023&11,933&1,081&918  \\
        \midrule
        Overall&119,555&8,440&3,813\\
        \bottomrule
    \end{tabular}
    \caption{The size of NHANES dataset}
    \label{tab:NHANES sample size}
\end{table}

\subsection{Data Preprocess}
(1) In examination variables, "systolic blood pressure" and "diastolic blood pressure" have three or four readings. Following the guidance from the American Heart Association(AHA), the final results of the systolic and diastolic blood pressure should be used as a reference for the average of the second and third times. This is to avoid the first time that the result is high because the patient is nervous, so that the calculation method can better reflect the true blood pressure level of the patient. AHA defines that Systolic exceed 130 mm Hg or diastolic exceed 80 mm Hg will be diagnosed as high blood pressure. Chinese hypertension guidelines also have a similar definition. We therefore calculated each respondent's blood pressure according to the rules.

(2) In Dietary variables, two days of dietary data were added after 2003 to help determine long-term eating habits. We used a paired t‑test to compare the mean dietary values of the same participant across two days. For example, in terms of protein intake, the t‑statistic was –1.33 and the p‑value was 0.183, which was well above the significance level. Therefore, from a statistical perspective, there was no significant difference in protein intake between the two days, and we could therefore use their average as the calorie intake for calculations. Note that 2009 didn't contain total sugar intake.

\begin{figure}[htbp]
  \centering
  \begin{subfigure}[b]{0.47\textwidth}
    \centering
    \includegraphics[width=\textwidth]{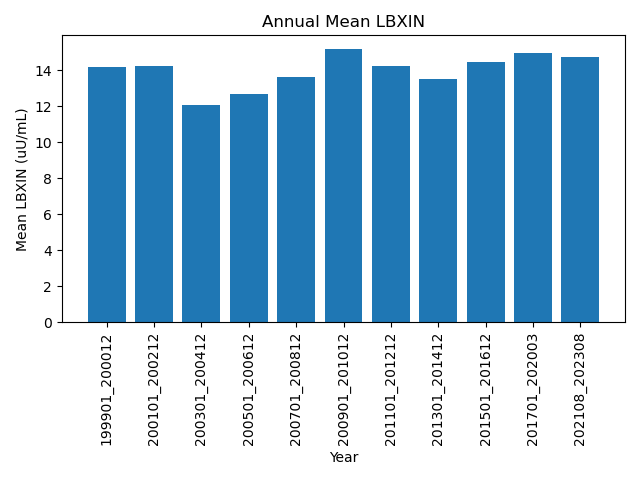}
    \caption{Insulin mean before adjustment}
    \label{fig:sub1}
  \end{subfigure}
  \quad
  \begin{subfigure}[b]{0.47\textwidth}
    \centering
    \includegraphics[width=\textwidth]{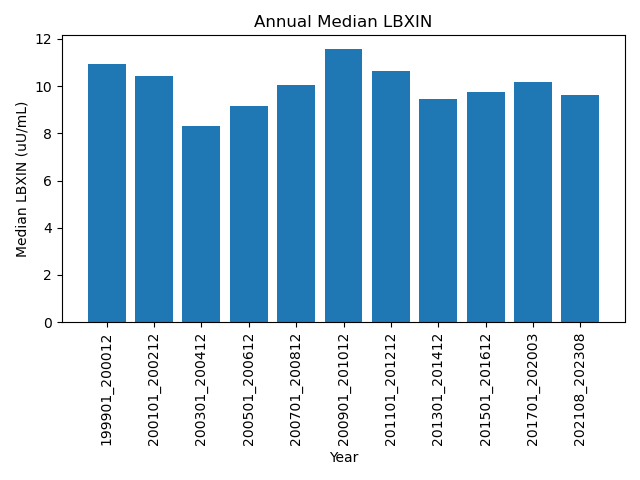}
    \caption{Insulin median before adjustment}
    \label{fig:sub2}
  \end{subfigure}

  \vspace{1em} 

  \begin{subfigure}[b]{0.47\textwidth}
    \centering
    \includegraphics[width=\textwidth]{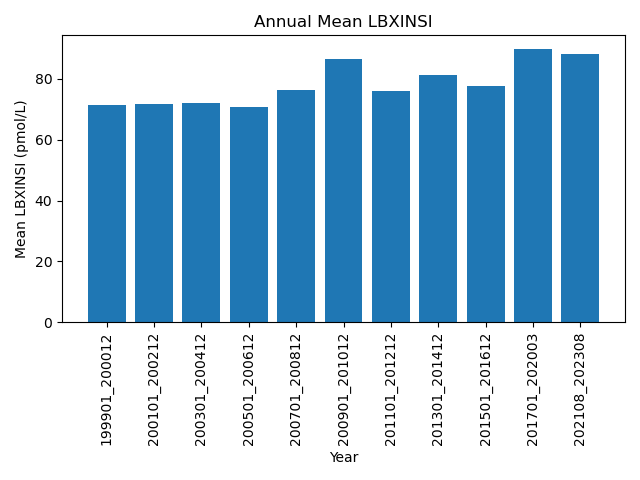}
    \caption{Insulin mean after adjustment}
    \label{fig:sub3}
  \end{subfigure}
  \quad
  \begin{subfigure}[b]{0.47\textwidth}
    \centering
    \includegraphics[width=\textwidth]{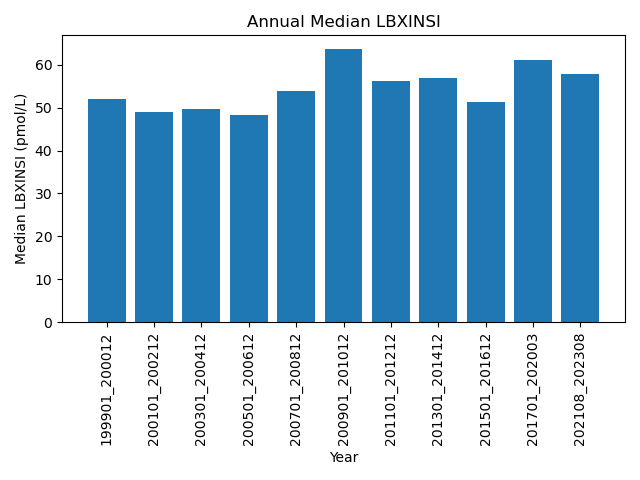}
    \caption{Insulin median after adjustment}
    \label{fig:sub4}
  \end{subfigure}

  \caption{Insulin mean and median before and after adjustment}
  \label{fig:insulin_mean_median_before_and_after_adjustment}
\end{figure}

(3) In Laboratory variables, the way insulin is assayed changes from year to year. In order to ensure the continuity of the data trend, the entire testing data from 1999-2023 was harmonized to the Tosoh AIA-PACK IRI based on the forward and backward regression equations given by the official NHANES. The unit also converted from $\mu$U/mL to pmol/L. Figure \ref{fig:insulin_mean_median_before_and_after_adjustment} shows the mean and median of insulin before and after adjustment.

\begin{figure}[htbp]
  \centering
  \begin{subfigure}[b]{0.47\textwidth}
    \centering
    \includegraphics[width=\textwidth]{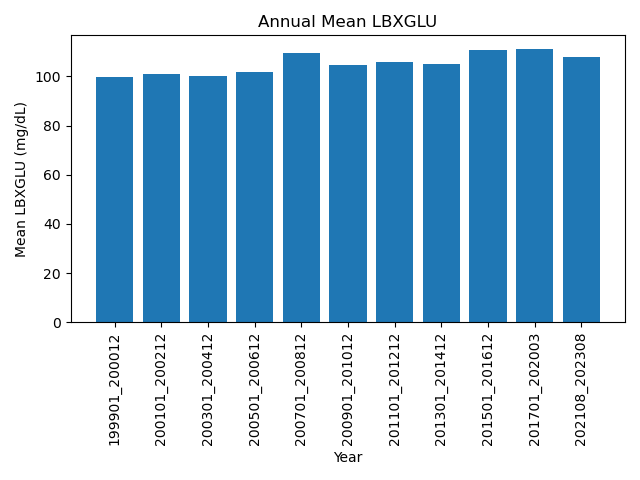}
    \caption{Fasting glucose mean before adjustment}
    \label{fig:sub1}
  \end{subfigure}
  \quad
  \begin{subfigure}[b]{0.47\textwidth}
    \centering
    \includegraphics[width=\textwidth]{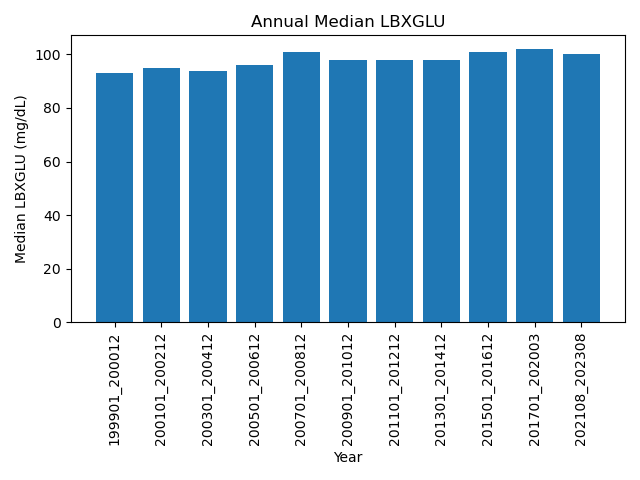}
    \caption{Fasting glucose median before adjustment}
    \label{fig:sub2}
  \end{subfigure}

  \vspace{1em} 

  \begin{subfigure}[b]{0.47\textwidth}
    \centering
    \includegraphics[width=\textwidth]{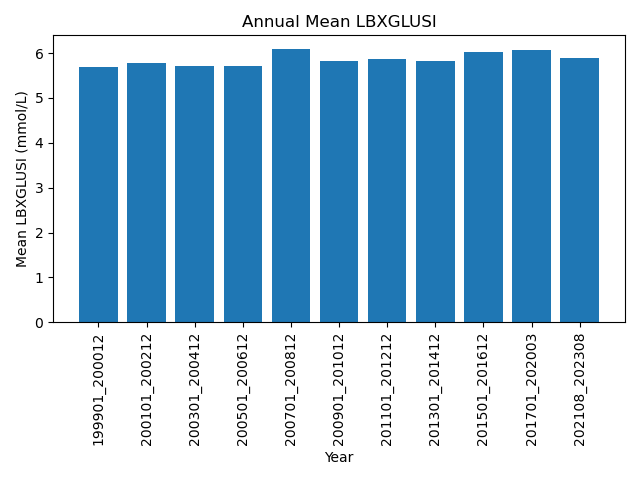}
    \caption{Fasting glucose mean after adjustment}
    \label{fig:sub3}
  \end{subfigure}
  \quad
  \begin{subfigure}[b]{0.47\textwidth}
    \centering
    \includegraphics[width=\textwidth]{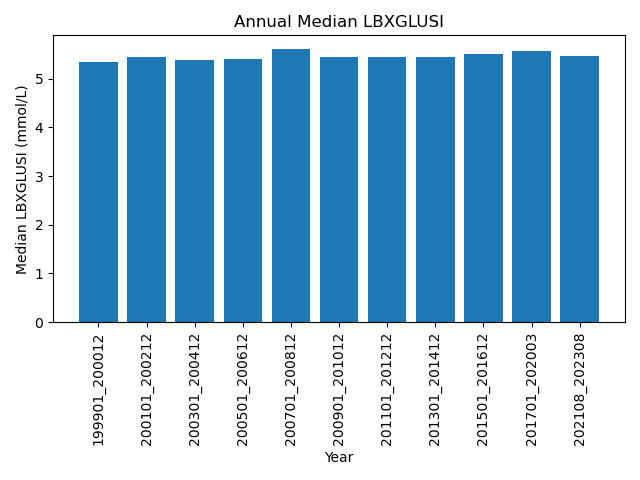}
    \caption{Fasting glucose median after adjustment}
    \label{fig:sub4}
  \end{subfigure}

  \caption{Fasting glucose mean and median before and after adjustment}
  \label{fig:fasting_glucose_mean_median_before_and_after_adjustment}
\end{figure}

We also adjusted fasting glucose measurements from different assay methods using the provided forward and backward regression equations. After adjustment, all data reflect values as measured by the Cobas C501 instrument. Figure \ref{fig:fasting_glucose_mean_median_before_and_after_adjustment} shows the mean and median of fasting glucose before and after adjuestment.

\begin{figure}[htbp]
  \centering
  \begin{subfigure}[b]{0.47\textwidth}
    \centering
    \includegraphics[width=\textwidth]{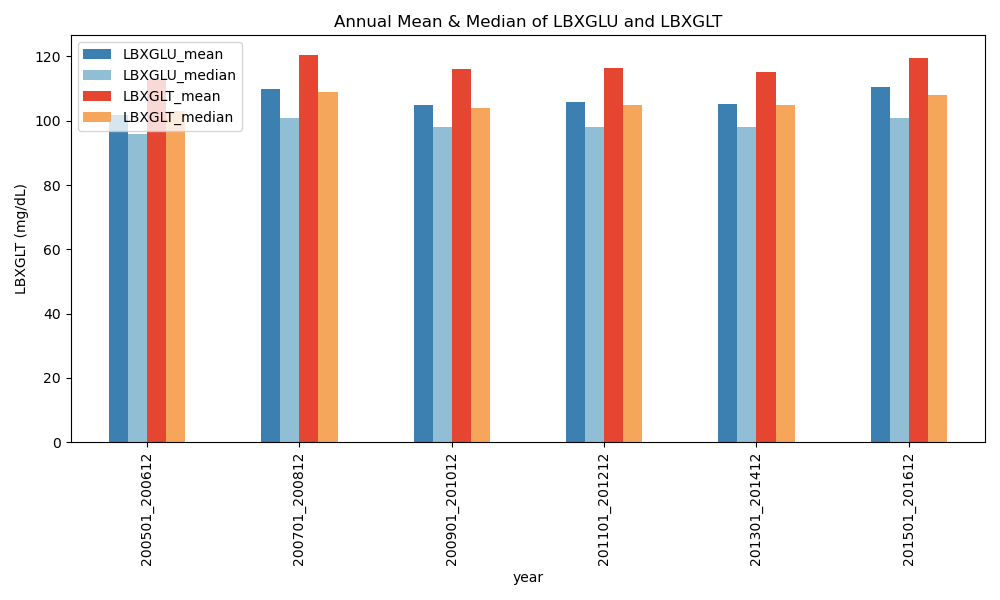}
    \caption{Fasting glucose and two-hour glucose's mean before adjustment}
    \label{fig:sub1}
  \end{subfigure}
  \quad
  \begin{subfigure}[b]{0.47\textwidth}
    \centering
    \includegraphics[width=\textwidth]{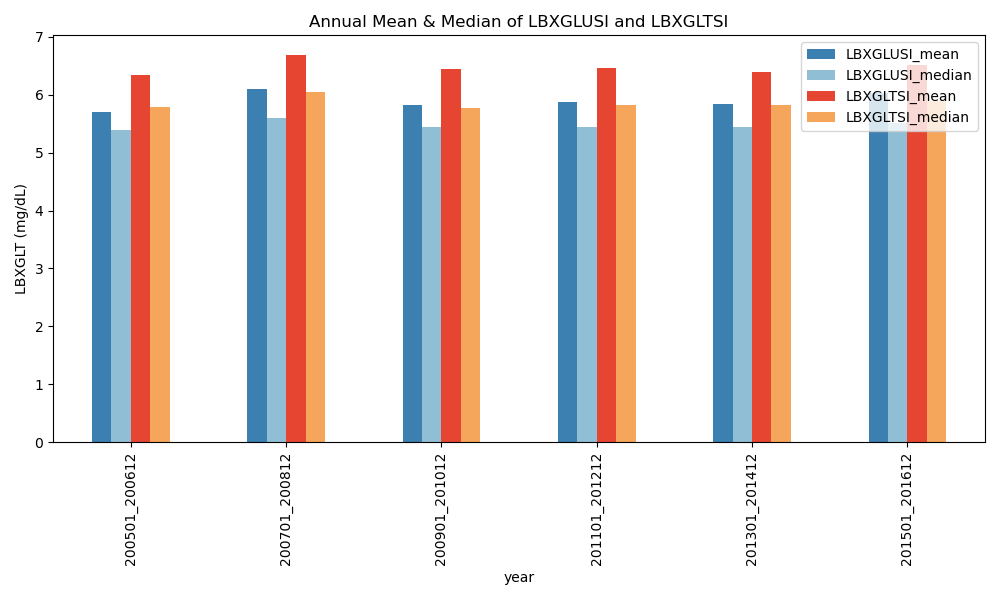}
    \caption{Fasting glucose and two-hour glucose's median after adjustment}
    \label{fig:sub2}
  \end{subfigure}

  \caption{Fasting glucose and two-hour glucose's mean and median before and after adjustment}
  \label{fig:fasting_and_two_hour_glucose_mean_median_before_and_after_adjustment}
\end{figure}

From 2005, NHANES started to include two-hour glucose in their records. However, they terminated it after 2017. The adjustment method were same with fasting glucose. Figure \ref{fig:fasting_and_two_hour_glucose_mean_median_before_and_after_adjustment} shows the comparison with fasting and two-hour glucose's mean and median before and after adjustment.

\begin{figure}[htbp]
  \centering
  \begin{subfigure}[b]{0.47\textwidth}
    \centering
    \includegraphics[width=\textwidth]{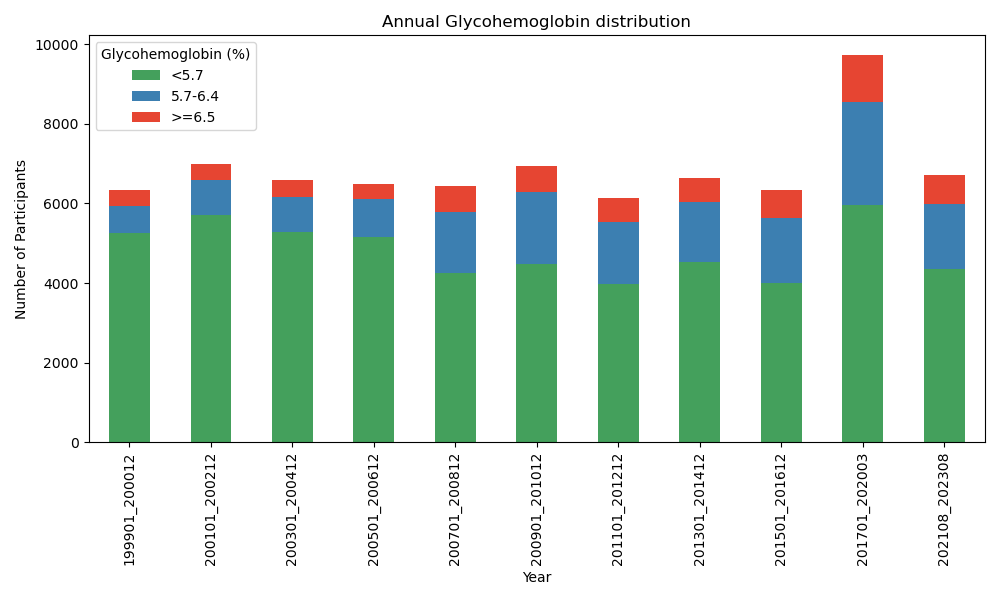}
  \end{subfigure}

  \caption{Annual Glycohemoglobin distribution of different levels}
  \label{fig:glycohemoglobin_distribution}
\end{figure}

The Clinical Practice Recommendations defines a glycohemoglobin level of 5.7\%-6.4\% as pre-diabetes and 6.5\% and above as diabetes. Figure \ref{fig:glycohemoglobin_distribution} shows the distribution of different levels of glycohemoglobin.

\begin{figure}[htbp]
  \centering
  \begin{subfigure}[b]{0.3\textwidth}
    \centering
    \includegraphics[width=\textwidth]{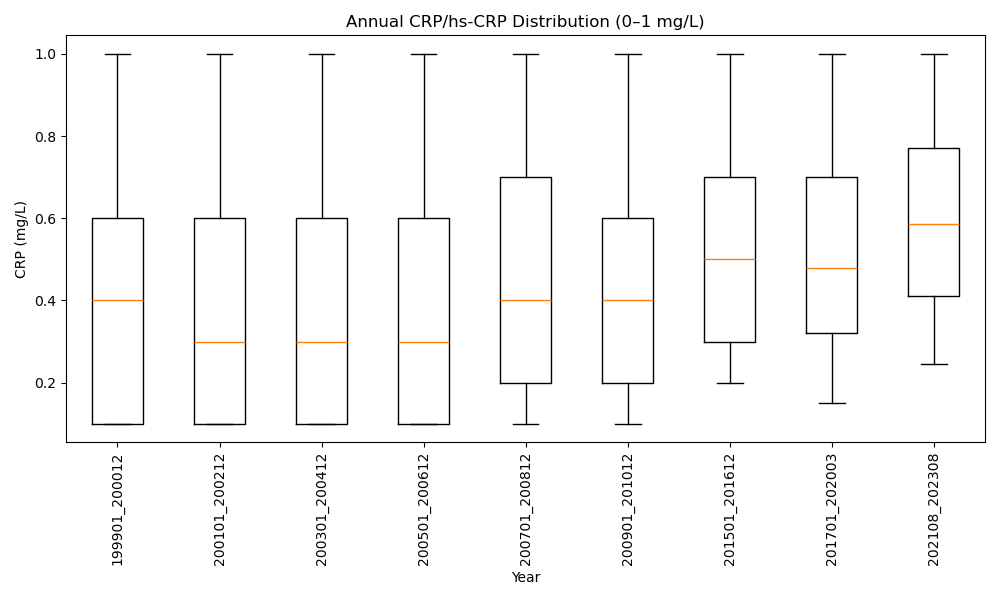}
    \caption{0 - 1 mg/dL}
    \label{fig:sub1}
  \end{subfigure}
  \quad
  \begin{subfigure}[b]{0.3\textwidth}
    \centering
    \includegraphics[width=\textwidth]{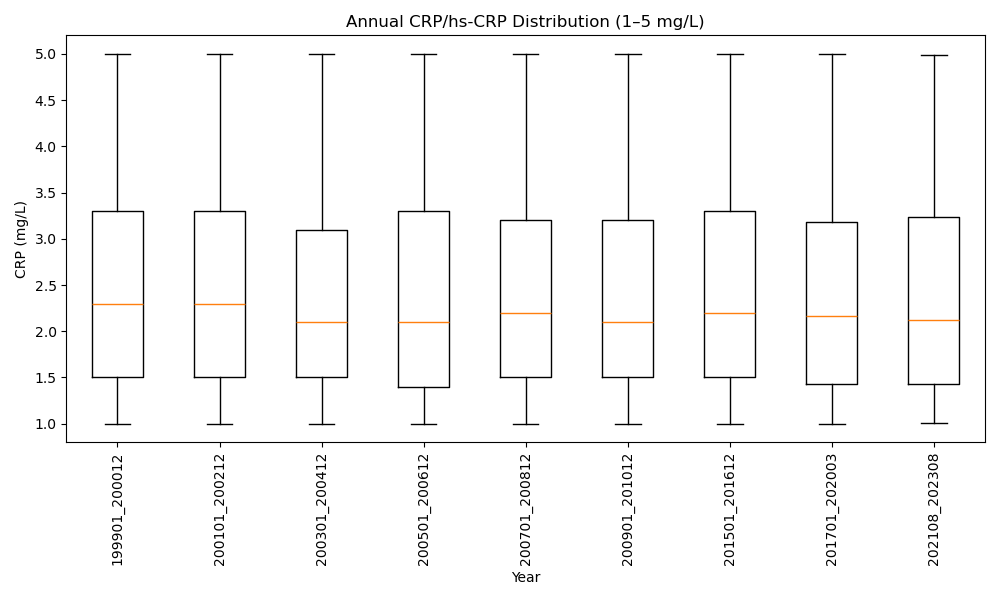}
    \caption{1 - 5 mg/dL}
    \label{fig:sub2}
  \end{subfigure}
  \quad
  \begin{subfigure}[b]{0.3\textwidth}
    \centering
    \includegraphics[width=\textwidth]{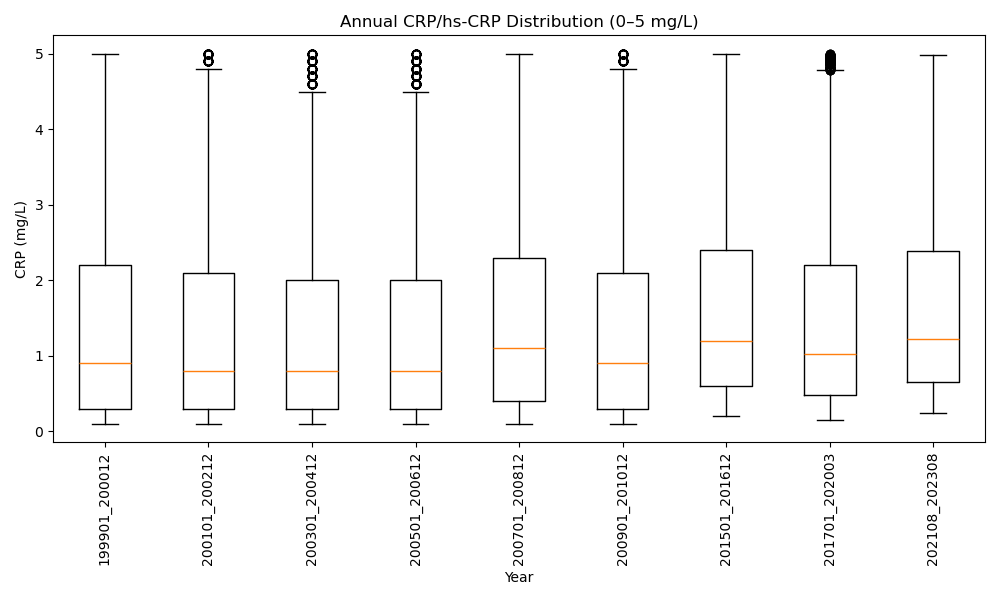}
    \caption{0 - 5 mg/dL}
    \label{fig:sub2}
  \end{subfigure}

  \caption{Annual distribution of C-reactive protein within three intervals}
  \label{fig:crp_distribution}
\end{figure}

The CRP data before 2009 refer to standard C-reactive protein (mg/dL), and after 2015, they refer to high-sensitivity C-reactive protein (mg/L). We converted the units to mg/L uniformly. After converting, Figure \ref{fig:crp_distribution} shows that the distribution of standard C-reactive protein and high-sensitivity C-reactive protein shares a similar pattern. Due to the assay detection limit, we removed 1849 records below the detection limit. Although CRP and hs-CRP tests measure the same protein, they cannot form a continuous trend due to different detection principles. Standard CRP measures general inflammation (healthy <0.9 mg/dL; 1–10 mg/dL mild infection; 10–50 mg/dL active disease; >50 mg/dL severe illness), whereas hs‑CRP is used for cardiovascular risk assessment (healthy <1 mg/L; 1–3 mg/L moderate risk; 3–10 mg/L high risk; >10 mg/L severe chronic conditions).

\begin{figure}[htbp]
  \centering
  \begin{subfigure}[b]{0.47\textwidth}
    \centering
    \includegraphics[width=\textwidth]{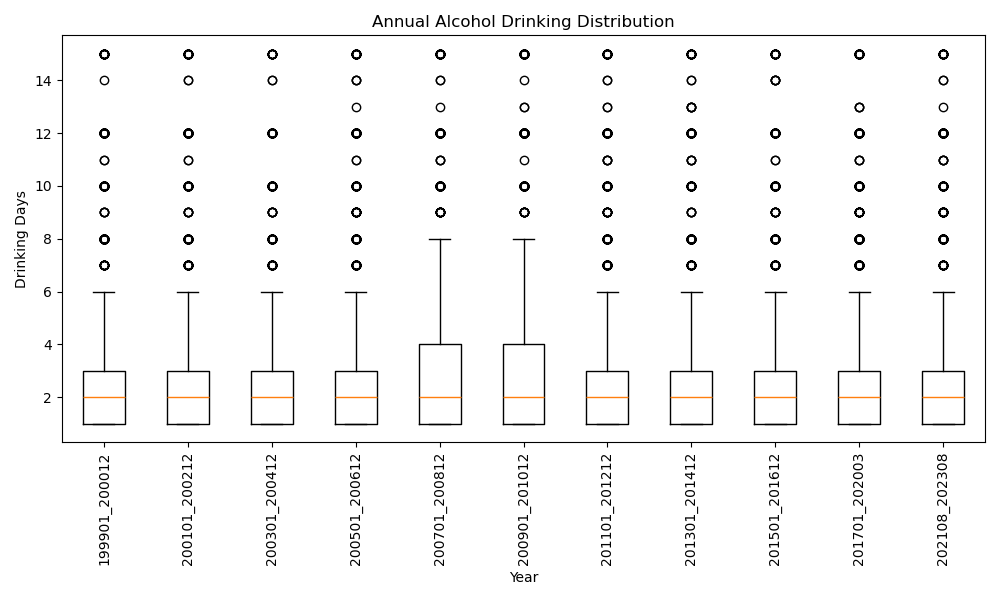}
  \end{subfigure}

  \caption{Annual alcohol drinking volume per day (cups)}
  \label{fig:alcohol_use_distribution}
\end{figure}

We used average alcohol drinking volume per day during the past 12 months as the measurement of alcohol use. We removed 60 invalid records from the dataset. More than 15 cups will be counted as 15 cups. Figure \ref{fig:alcohol_use_distribution} shows the distribution of everyday alcohol drinking volume in the past 12 months. We can see over the past 20 years, participants in this question have drank an average of two alcoholic drinks per day.

\begin{figure}[htbp]
  \centering
  \begin{subfigure}[b]{0.47\textwidth}
    \centering
    \includegraphics[width=\textwidth]{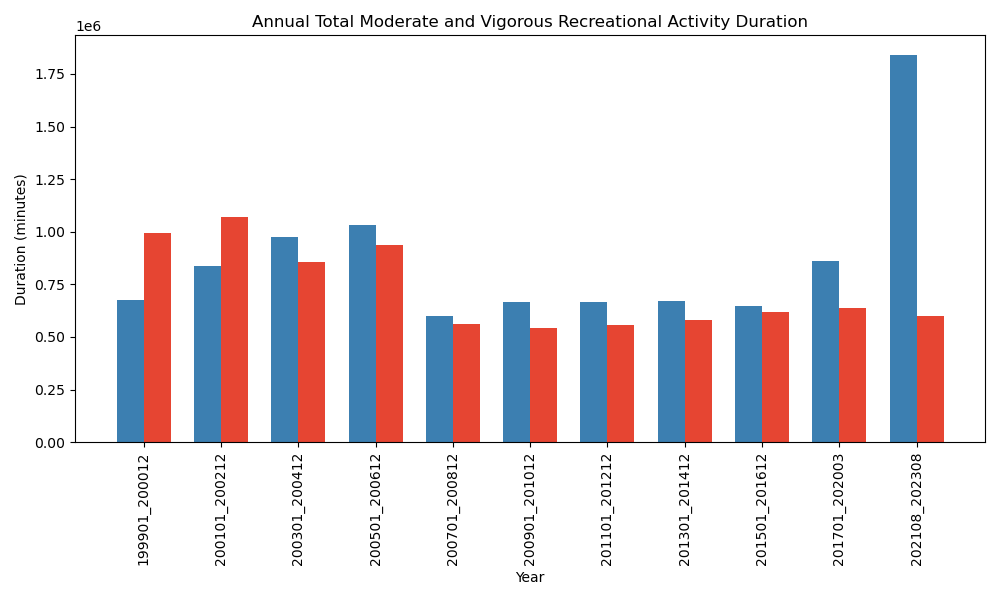}
    \caption{Total duration}
    \label{fig:sub1}
  \end{subfigure}
  \quad
  \begin{subfigure}[b]{0.47\textwidth}
    \centering
    \includegraphics[width=\textwidth]{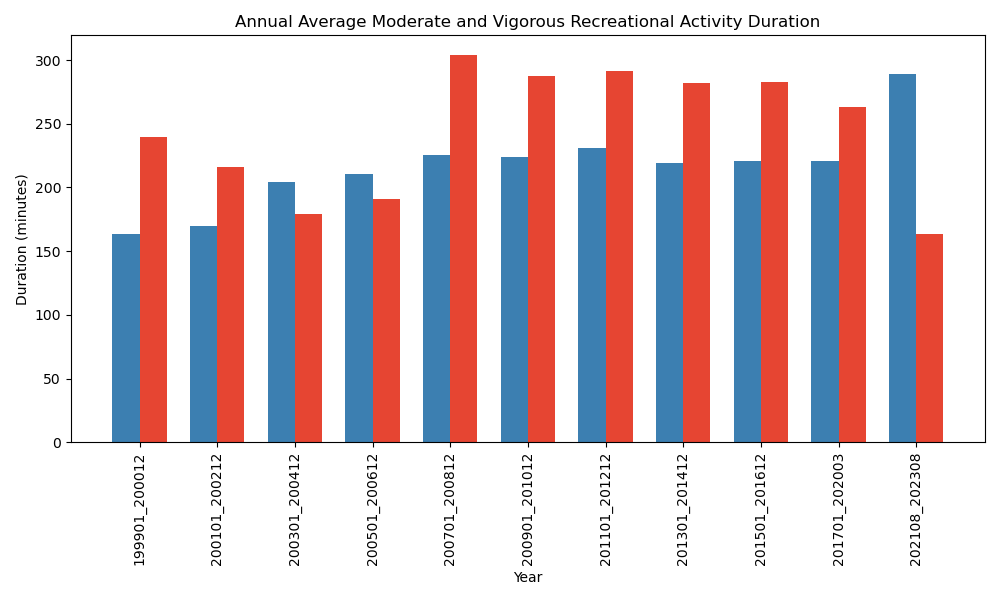}
    \caption{Average duration}
    \label{fig:sub2}
  \end{subfigure}

  \caption{Moderate and vigorous intensity recreational activity duration (minutes/week)}
  \label{fig:physical_activity_distribution}
\end{figure}

The physical activity data include household, work, and recreational activities, each classified as moderate or vigorous intensity. Because diabetes prevention and management rely on extra exercise, we focused exclusively on recreational activity. Although the 1999–2006 surveys listed detailed exercise types, later cycles did not; therefore, we aggregated all recreational activities by intensity (moderate and vigorous) and standardized the units to minutes per week. Figure \ref{fig:physical_activity_distribution} shows the duration of moderate and vigorous intensity recreational activity duration for each year.

\begin{figure}[htbp]
  \centering
  \begin{subfigure}[b]{0.47\textwidth}
    \centering
    \includegraphics[width=\textwidth]{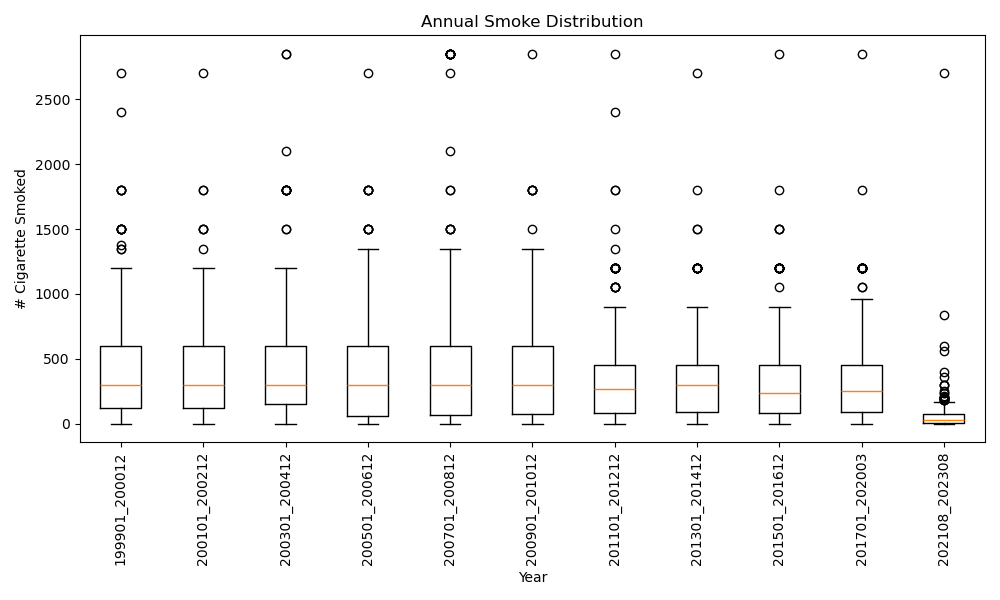}
  \end{subfigure}

  \caption{Annual cigarette consumption over a 30-day period}
  \label{fig:smoke_distribution}
\end{figure}

As for smoking behavior, we use the equation \ref{smoking_equation} to calculate the average number of cigarettes smoked per month. Figure \ref{fig:smoke_distribution} shows the distribution of the number of cigarettes smoked during a 30-day period. According to the figure, participants involved in this question smoked an average of 300 cigarettes in the last 30 days. Participants who answered with a value greater than 95 were uniformly considered to be 95.

\begin{equation}
    \text{SMD}=\text{SMD}_{\text{\# days smoked}} \times \text{SMD}_{\text{\# cigarettes/day}}
\label{smoking_equation}
\end{equation}

\begin{figure}[htbp]
  \centering
  \begin{subfigure}[b]{0.47\textwidth}
    \centering
    \includegraphics[width=\textwidth]{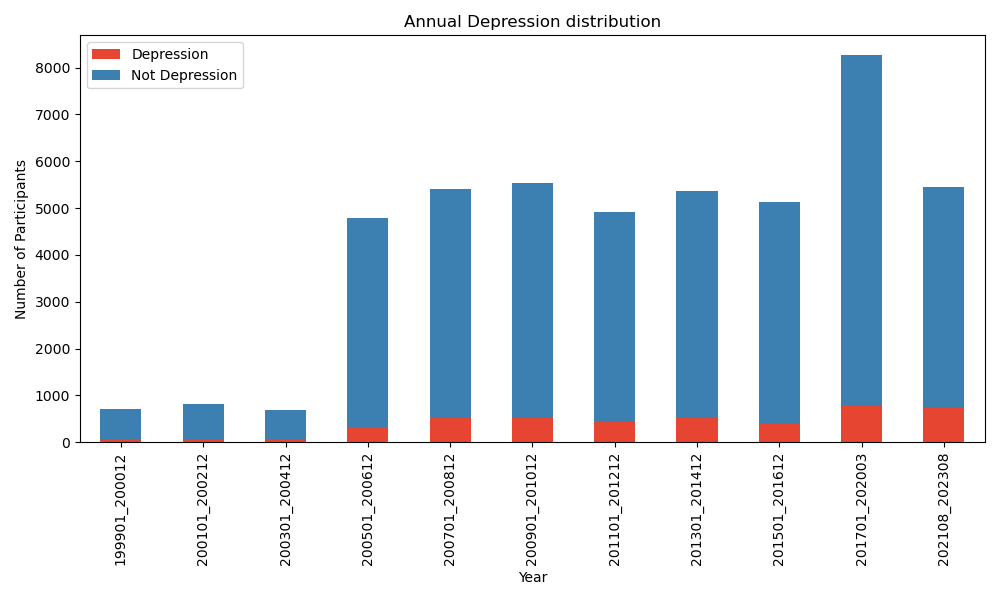}
  \end{subfigure}

  \caption{Annual depression distribution}
  \label{fig:depression_distribution}
\end{figure}

From 1999 to 2004, NHANES used CIDI-Auto 2.1 questionnaire to assess mental health. We used the score given to classify depression. Starting in 2005, a depression screener, Patient Health Questionnaire (PHQ), has been used to detect depression. The PHQ-9 total score of 10 or above is generally used as the cutoff indicative of clinically significant depression. Therefore, we summed the scores of 9 questions, and participants with a total score of 10 or higher are classified as having depression. Figure \ref{fig:depression_distribution} shows 

\begin{figure}[htbp]
  \centering
  \begin{subfigure}[b]{0.47\textwidth}
    \centering
    \includegraphics[width=\textwidth]{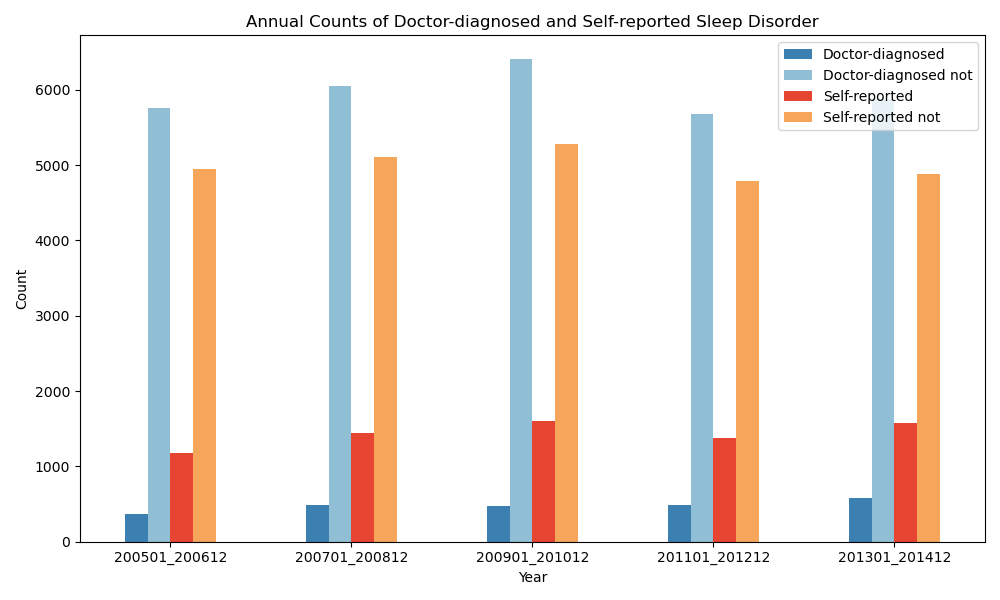}
  \end{subfigure}

  \caption{Annual counts of doctor-diagnosed and self-reported sleep disorder}
  \label{fig:SLD_self_doctor}
\end{figure}

\begin{figure}[htbp]
  \centering
  \begin{subfigure}[b]{0.47\textwidth}
    \centering
    \includegraphics[width=\textwidth]{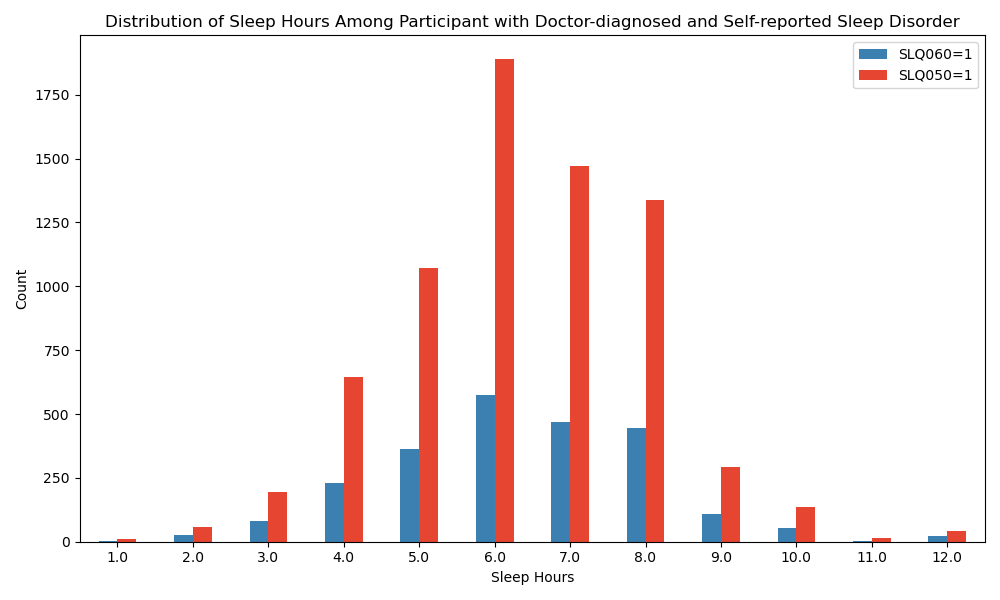}
  \end{subfigure}

  \caption{Distribution of sleep hours among participant with doctor-diagnosed and self-reported sleep disorder}
  \label{fig:SLD_distribution_positive}
\end{figure}

Because the sleep disorder questionnaire items have changed over the years, no single variable can be used for continuous analysis. Therefore, we defined a Sleep Disorder Indicator (SDI) that uses the variables available in each questionnaire and applies Equation \ref{sleep_disorder_equation} to determine the presence of a sleep disorder. We analyzed data from 2005 to 2013 that included sleep duration, doctor‑diagnosed sleep disorders, and self‑reported sleep disorder. As Figure \ref{fig:SLD_self_doctor} shows, self‑reported records far outnumber doctor-diagnosed, suggesting many participants never sought medical evaluation. Figure \ref{fig:SLD_distribution_positive} reveals that, regardless of doctor-diagnosed or self-reported, those identified with a sleep disorder most often slept 5–8 hours per night. Normally, any deviation besides 7–9 hours is considered a disorder. The overlap prevents our SDI from distinguishing affected individuals. Consequently, we defined sleep disorder solely by self‑reported sleep problems.

\begin{figure}[htbp]
  \centering
  \begin{subfigure}[b]{0.47\textwidth}
    \centering
    \includegraphics[width=\textwidth]{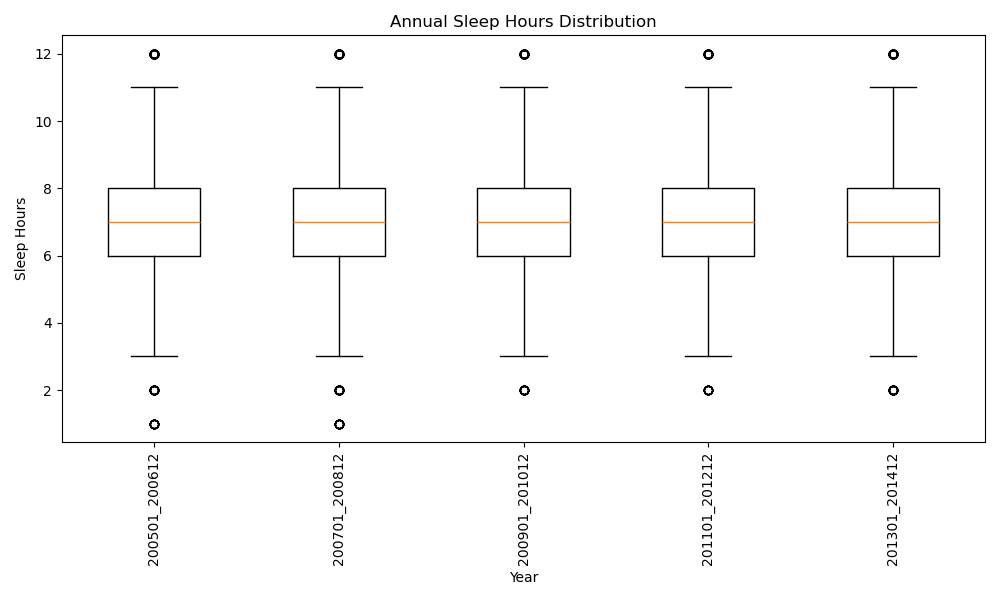}
  \end{subfigure}

  \caption{Annual sleep disorder distribution}
  \label{fig:SLD_distribution}
\end{figure}

\begin{equation}
\label{eq:sleep_disorder}
\begin{aligned}
\mathrm{SDI}_i &=
\begin{cases}
1, & \text{if }
    \begin{aligned}[t]
    &(D_i<7 \;or\; D_i>9) \\
    &or\;\mathrm{SLQ060}_i=1 \\
    &or\;\mathrm{SLQ050}_i=1 \\
    &or\;\mathrm{SLQ120}_i\ge3 \\
    \end{aligned}
\\
0, & \text{otherwise}
\end{cases}
\\[1ex]
\text{where,}
\\
&D_i = \frac{5\,\mathrm{SLD012}_i + 2\,\mathrm{SLD013}_i}{7} \quad\text{is the average sleep hours everyday},\\
&\mathrm{SLD012}_i\quad\text{is the sleep hours on weekdays},\\
&\mathrm{SLD013}_i\quad\text{is the sleep hours on weekends},\\
&\mathrm{SLQ060}_i = 1\quad\text{if doctor‐diagnosed sleep disorder},\\
&\mathrm{SLQ050}_i = 1\quad\text{if self‐reported trouble sleeping},\\
&\mathrm{SLQ120}_i = \text{frequency of daytime sleepiness (0–4)}.
\end{aligned}
\end{equation}

\begin{figure}[htbp]
  \centering
  \begin{subfigure}[b]{\textwidth}
    \centering
    \includegraphics[width=\textwidth]{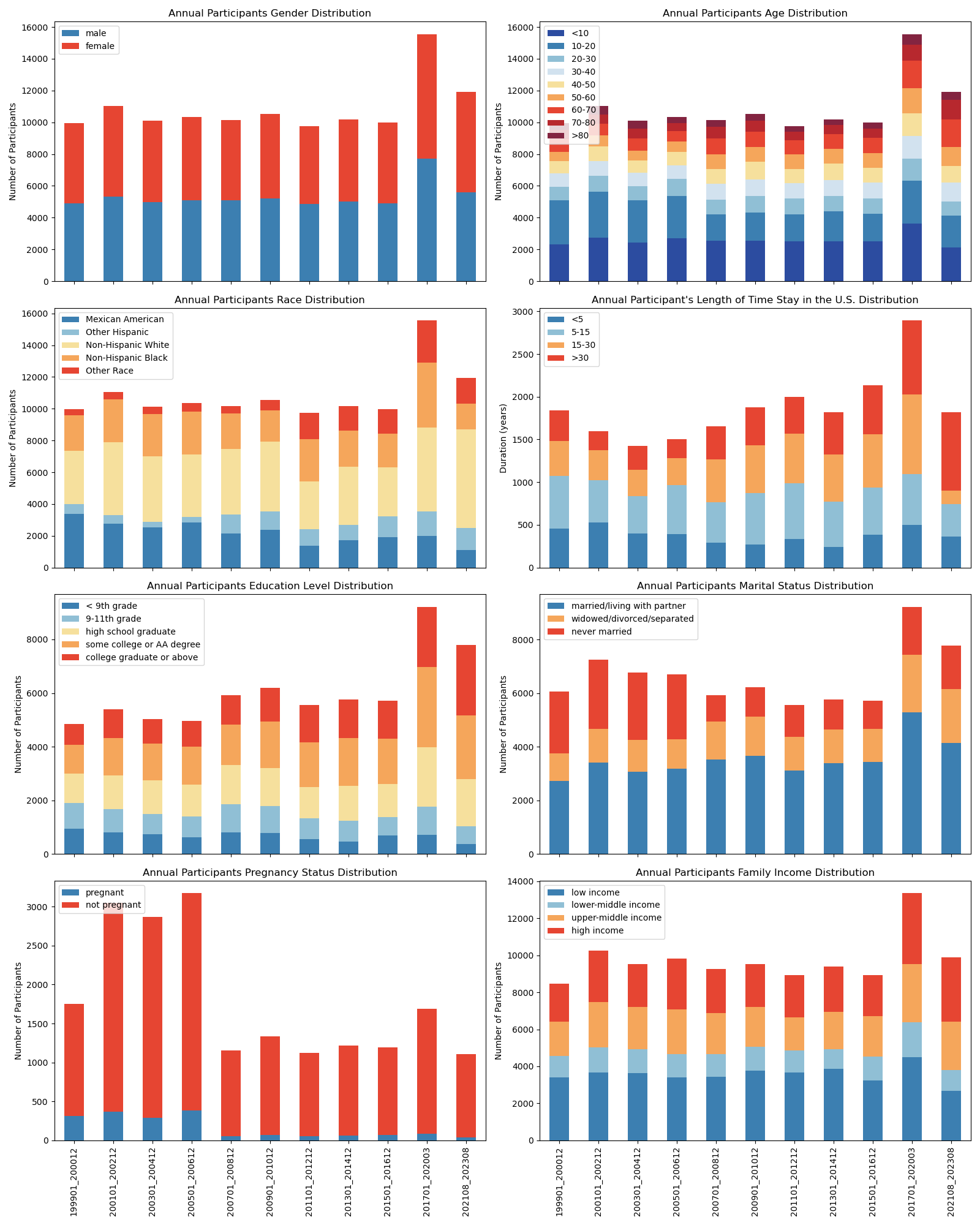}
  \end{subfigure}

  \caption{Distribution of demographics data}
  \label{fig:DEMO_distribution}
\end{figure}

Demographics variables contain 8 variables that describe participants' features, including gender, age, race, time stay in the U.S., pregnancy status, education level, marital status, and ratio of family income to poverty. Figure \ref{fig:DEMO_distribution} shows the distribution of each demographic's variables.

Table \ref{tab:NHANES sample size} shows the number of records of the five categories of data from 1999 to 2023. Demographics data includes gender, age, race, marital status, time stay in the U.S., education level, pregnancy status, and family poverty income ratio. Dietary data includes daily energy, protein, carbohydrate, dietary fiber, fat, cholesterol, and sugar intake. Examination data includes body mass index, blood pressure. Laboratory data includes glucose, insulin, C-reactive protein, and glycohemoglobin. Questionnaire data includes alcohol drinking behavior, physical activity, smoking behavior, depression, and sleep disorder. All responses that are marked as ``don't know" or ``reject" has been removed from the dataset.

\begin{sidewaystable}[htbp]
    \centering
    \begin{tabular}{lccccccccccccccc}
        \toprule
        \textbf{Year} & \textbf{BPX} & \textbf{BMI} & \textbf{Dietary} & \textbf{LBXGLUSI} & \textbf{OGTT} & \textbf{LBXINSI} & \textbf{LBXGH} & \textbf{CRP} & \textbf{ALQ} & \textbf{SMD} & \textbf{CIQ} & \textbf{SLQ} & \textbf{PAQ} & \textbf{DEMO} & \textbf{DEMO2}\\
        \midrule
        1999-2000&5,613&8,462&8,725&3,035&-&2,986&6,343&7,493&2,650&997&714&-&4,137&335&968\\
        2001-2002&6,160&9,010&9,701&3,441&-&3,379&6,999&8,410&2,977&1,171&812&-&4,952&335&980\\
        2003-2004&4,852&8,687&8,213&3,169&-&3,136&6,601&8,254&2,696&1,117&691&-&4,782&287&946\\
        2005-2006&5,551&8,949&8,257&3,128&2,319&3,080&6,493&8,172&2,820&1,385&4,799&6,130&4,901&375&993\\
        2007-2008&6,678&8,861&7,711&3,110&2,233&3,071&6,427&6,809&3,312&1,469&5,415&6,540&3,403&249&1,238\\
        2009-2010&7,438&9,412&8,279&3,386&2,574&3,337&6,930&8,299&3,664&1,504&5,546&6,886&3,736&297&1,396\\
        2011-2012&6,797&8,602&7,486&3,033&2,287&2,881&6,145&-&3,329&1,216&4,925&6,172&3,628&317&1,485\\
        2013-2014&7,300&9,055&7,449&3,172&2,345&3,093&6,643&-&3,593&1,246&5,372&6,462&3,851&327&1,409\\
        2015-2016&7,231&8,756&6,863&2,972&2,084&2,921&6,326&6,602&3,375&1,107&5,134&6,323&3,687&336&1,618\\
        2017-2020&10,285&13,137&10,610&4,744&-&4,625&9,737&11,216&5,853&1,680&8,276&10,187&4,606&374&2,058\\
        2021-2023&7,479&8,471&5,828&3,672&-&3,510&6,715&7,096&4,055&242&5,455&-&6,475&197&1,188\\
        \midrule
        Overall&75,384&101,402&89,122&36,862&13,842&32,682&75,359&72,351&38,324&13,134&47,139&48,700&48,158&3,429&14,279\\
        \bottomrule
    \multicolumn{16}{p{\textwidth}}
    {Notes: 1999-2000 doesn't have sugar intake; BPX refers to blood pressure; BMI refers to body mass index(kg/$m^2$); LBXGLUSI refers to Fasting Plasma Glucose(mmol/L); OGTT refers to Two-hour Glucose(mmol/L); LBXINSI refers to Insulin(pmol/L); LBXGH refers to Glycohemoglobin(\%); CRP refers to C-reactive protein(mg/L); ALQ refers to alcohol use; SMD refers to smoke (\# cigarettes/month); CIQ refers to depression; SLQ refers to sleep disorder; PAQ refers to physical activity (minutes/week); DEMO refers to 8 demographics data; DEMO2 refers to 7 demographics data without pregnancy.}
    \end{tabular}
    \caption{The size of NHANES dataset}
    \label{tab:NHANES sample size}
\end{sidewaystable}

\section{Model}

\begin{figure}[htbp]
  \centering
  \begin{subfigure}[b]{\textwidth}
    \centering
    \includegraphics[width=\textwidth]{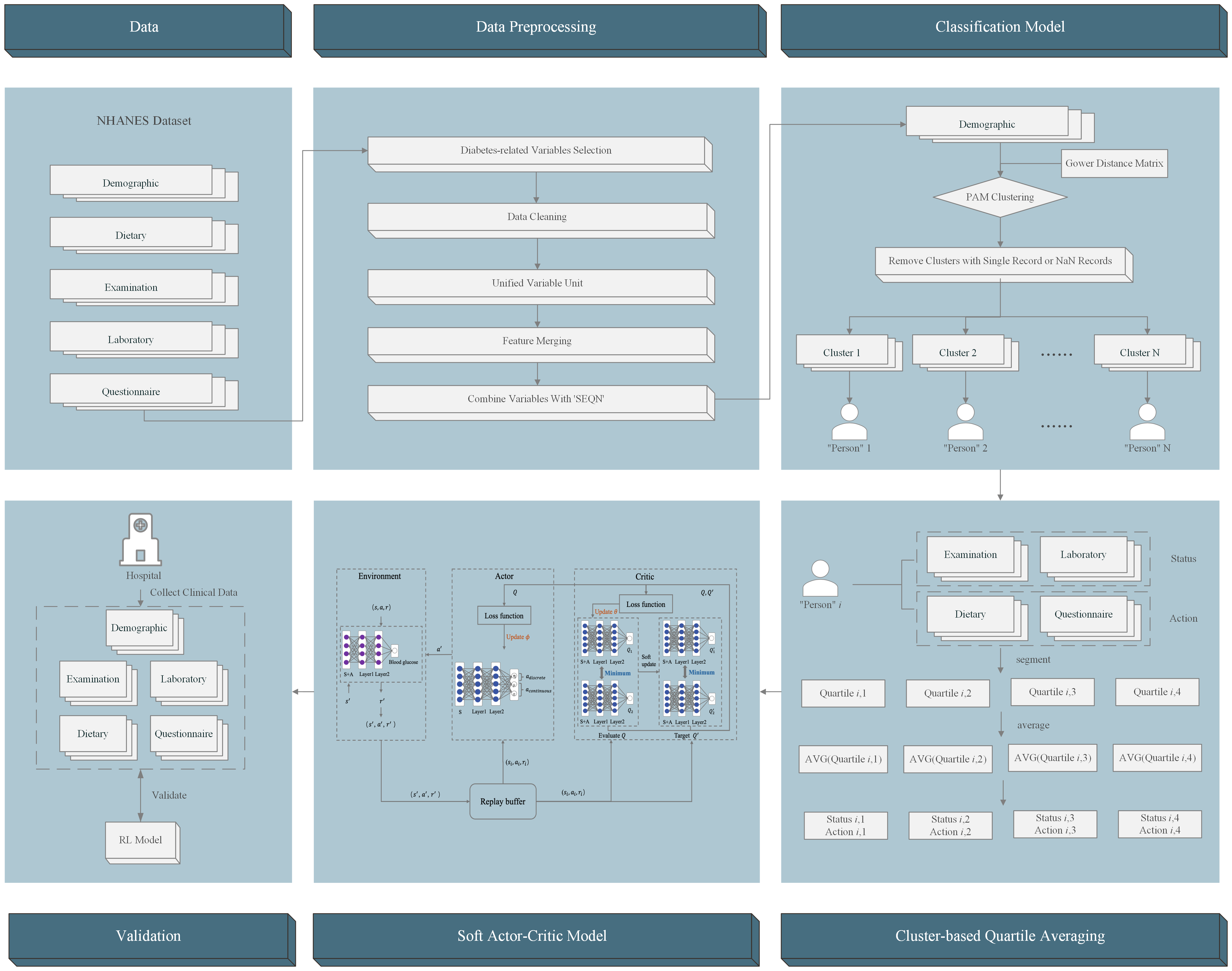}
  \end{subfigure}

  \caption{Project Workflow}
  \label{fig:workflow}
\end{figure}

Our model has two parts. The first part is a classification model that segments NHANES participants by demographic variables. The second part is a reinforcement learning model that aggregates those segments and learns how individuals with different demographic profiles respond to various lifestyle habits, with the goal of minimizing diabetes risk or achieving remission. Figure \ref{fig:workflow} shows the workflow of the project.

\subsection{Unsupervised Classification Model}
Because NHANES does not track the same individuals over time, we created synthetic “individuals” by grouping records with similar demographic profiles. We classified each record using gender, age, race, marital status, length of U.S. residence, education level, and family poverty‑income ratio. We excluded pregnancy status at this stage, as it applies only to female participants and would greatly reduce the sample size. 

Demographic data include both numerical (age and family poverty‑income ratio) and categorical types (gender, race, marital status, length of U.S. residence, and education level). Common classification algorithms such as random forest and support vector machine cannot directly handle mixed data with numerical and categorical variables. Instead, they require one-hot encoding processing before input. Therefore, we chose the PAM algorithm to classify participants with different demographic features.

Firstly, we calculated the Gower distance matrix for 7 features. The Gower distance can measure the dissimilarity between two records. For a pair of records $i$ and $j$, its Gower distance $d^g(i,j)$ is calculated as equation \ref{eq:gower_distance}:

\begin{equation}
\label{eq:gower_distance}
\begin{aligned}
&d^g(i, j) = \frac{\sum^p_{s=1} w_s\delta^{(s)}_{ij}}{\sum^p_{s=1}w_s}
\\[1ex]
\text{where,}
\\
&p\text{ is the total number of variables},\\
&\delta^{(s)}_{ij}\text{ is the dissimilarity of variable } i \text{ and }j,\\
&w_s\text{ is the weight for variables},\\
&\delta^{(s)}_{ij}\in[0,1],\\
&d^g(i, j)\in[0,1],\\
&i\in\{1, 2, ..., N\}, j\in\{1, 2, ..., N\},\\
&\text{N is the total number of records}.
\end{aligned}
\end{equation}

The dissimilarity $\delta^{(s)}_{ij}$ is defined as equation \ref{eq:gower_distance_detail}:

\begin{equation}
\label{eq:gower_distance_detail}
\begin{aligned}
\delta^{(s)}_{ij} &=
\begin{cases}
\frac{|x_{is} - x_{js}|}{R_s}, &s \text{ is numeric variable}, \\
1\{x_{is}\neq x_{js}\}, &s \text{ is categorical variable}.\\
\end{cases}
\\[1ex]
\text{where,}
\\
&R_s = \max_{k} x_{ks} - \min_{k} x_{ks} \text{ is the observed range of variable} s,\\
&x_{ks} \text{ is the value of variable } s \text{ on the } k^{th} \text{record},\\
&x_{is} \text{ is the value of variable } s \text{ on the } i^{th} \text{record},\\
&x_{js} \text{ is the value of variable } s \text{ on the } j^{th} \text{record},\\
&1\{\cdot\} \text{ is the indicator function, 1 if the condition is matched and 0 otherwise}\\
&i\in\{1, 2, ..., N\}, j\in\{1, 2, ..., N\},\\
&N\text{ is the total number of records}.
\end{aligned}
\end{equation}

Then, we applied the partitioning around medoids (PAM) algorithm to cluster participants. The PAM algorithm, similar to the K-means algorithm, is a clustering algorithm, whose mediods have minimal average dissimilarity to all other points in that cluster. The dissimilarity score is the Gower distance matrix that we calculated above. The objective function of PAM is shown as equation \ref{eq:PAM_object}:

\begin{equation}
\label{eq:PAM_object}
\begin{aligned}
&\sum_{i=1}^N \min_{m \in M} d^g(i, m)
\\[1ex]
\text{where,}
\\
&M \text{ is the set of medoids},\\
&d^g(i, m) \text{is the dissimilarity score between point }i\text{ and medoid }m,\\
&N\text{ is the total number of records}.
\end{aligned}
\end{equation}

\begin{figure}[htbp]
  \centering
  \begin{subfigure}[b]{0.47\textwidth}
    \centering
    \includegraphics[width=\textwidth]{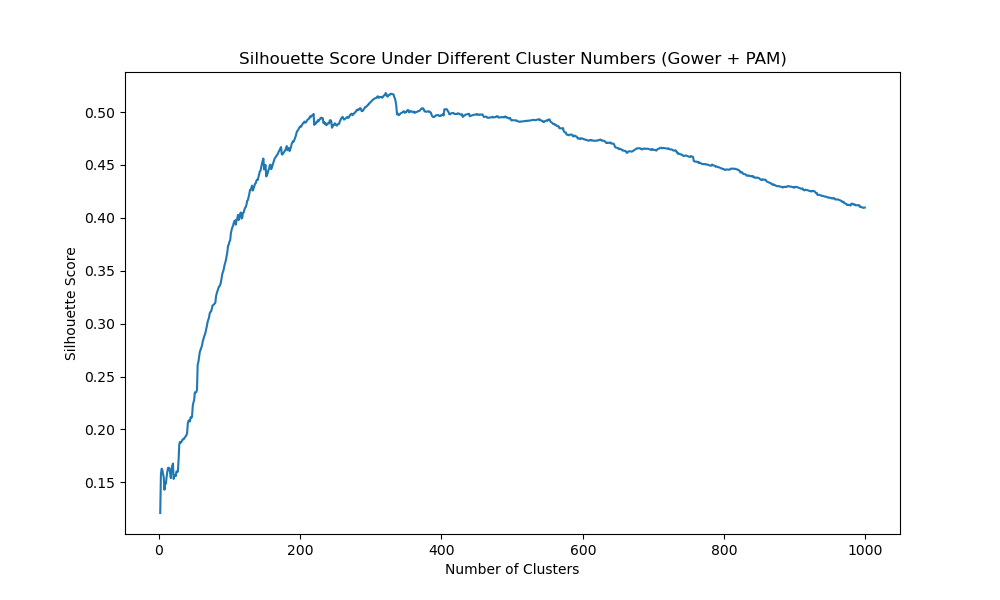}
    \caption{With Pregnancy Status \& Remove NaN Value (Best k = 321)}
    \label{fig:sub1}
  \end{subfigure}
  \quad
  \begin{subfigure}[b]{0.47\textwidth}
    \centering
    \includegraphics[width=\textwidth]{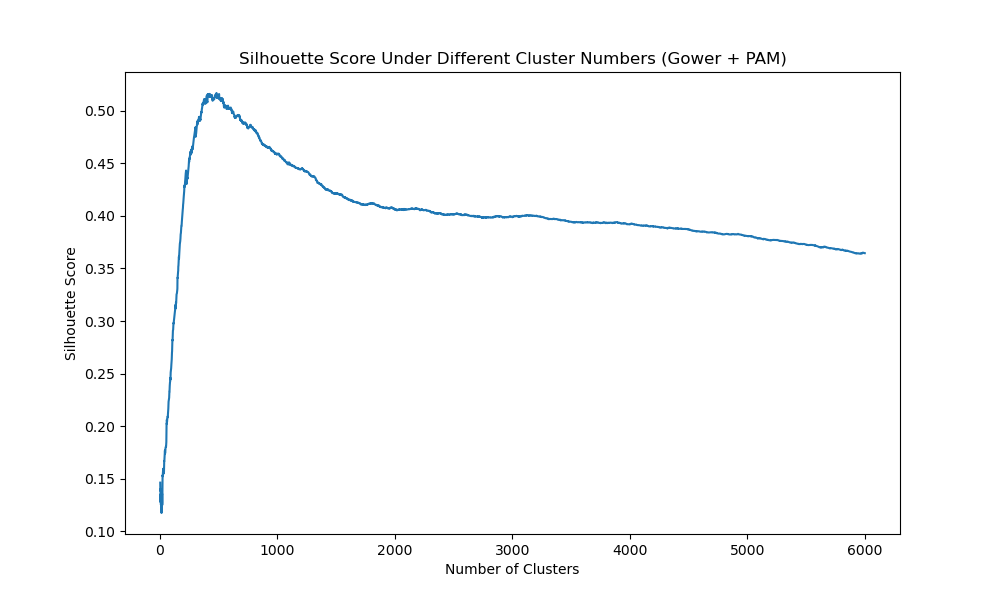}
    \caption{Without Pregnancy Status \& Remove NaN Value (Best k = 481)}
    \label{fig:sub2}
  \end{subfigure}
  \vspace{1em}
  \begin{subfigure}[b]{0.47\textwidth}
    \centering
    \includegraphics[width=\textwidth]{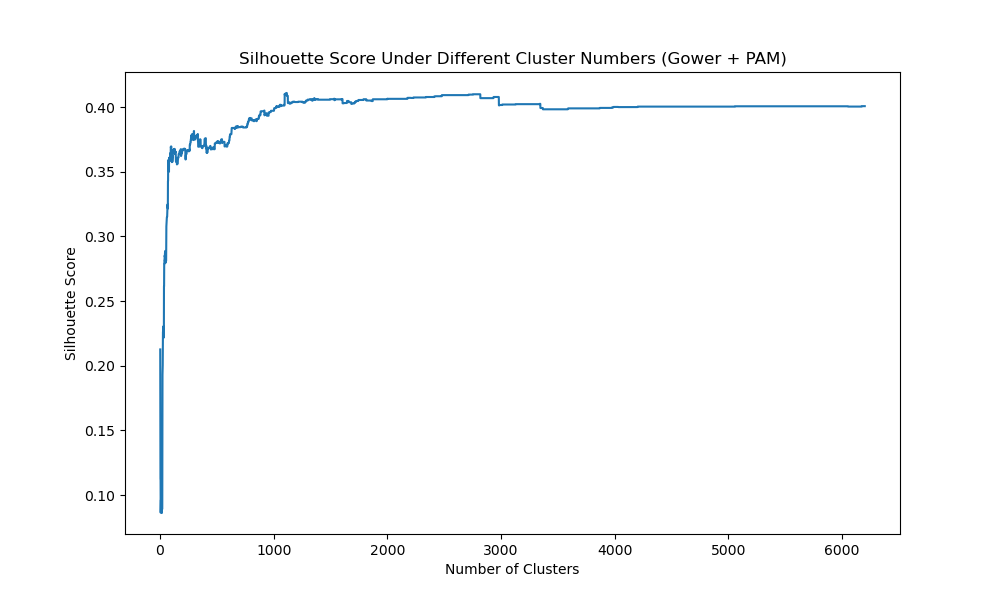}
    \caption{Without Pregnancy Status \& Not Remove NaN Value (Best k = 1,109)}
    \label{fig:sub3}
  \end{subfigure}
  \quad
  \begin{subfigure}[b]{0.47\textwidth}
    \centering
    \includegraphics[width=\textwidth]{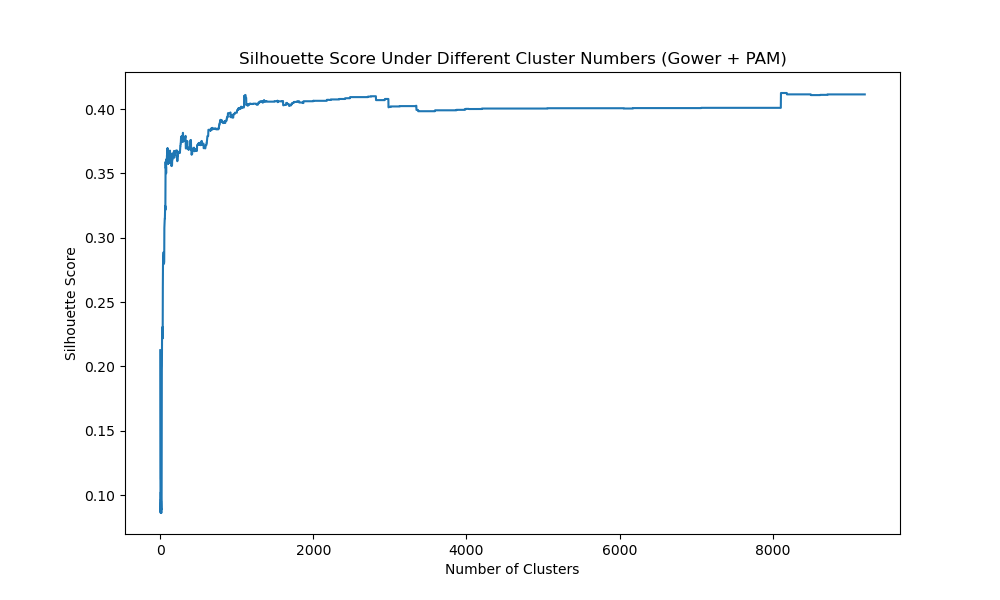}
    \caption{Without Pregnancy Status \& Not Remove NaN Value (Best k = 8,104)}
    \label{fig:sub3}
  \end{subfigure}
  
  \caption{Silhouette Score}
  \label{fig:silhouette_score}
\end{figure}

We ran clustering for k from 2 to 6,000 and used the silhouette score to determine the optimal number of clusters. The silhouette score measures how well each point assigned to its cluster. Figure \ref{fig:silhouette_score} shows the silhouette score under different number of clusters. Removing both NaNs and pregnancy status yields an optimal k=481 with a silhouette score of 0.517 (a total of 14,279 records), and removing only pregnancy status yields k=1,109 with a silhouette score of 0.411 (a total of 119,555 records). Although the silhouette score peaked at 0.4125 for k = 8104 when removing only pregnancy status, we discarded it because it yielded too many clusters.

\begin{figure}[htbp]
  \centering
  \begin{subfigure}[b]{0.47\textwidth}
    \centering
    \includegraphics[width=\textwidth]{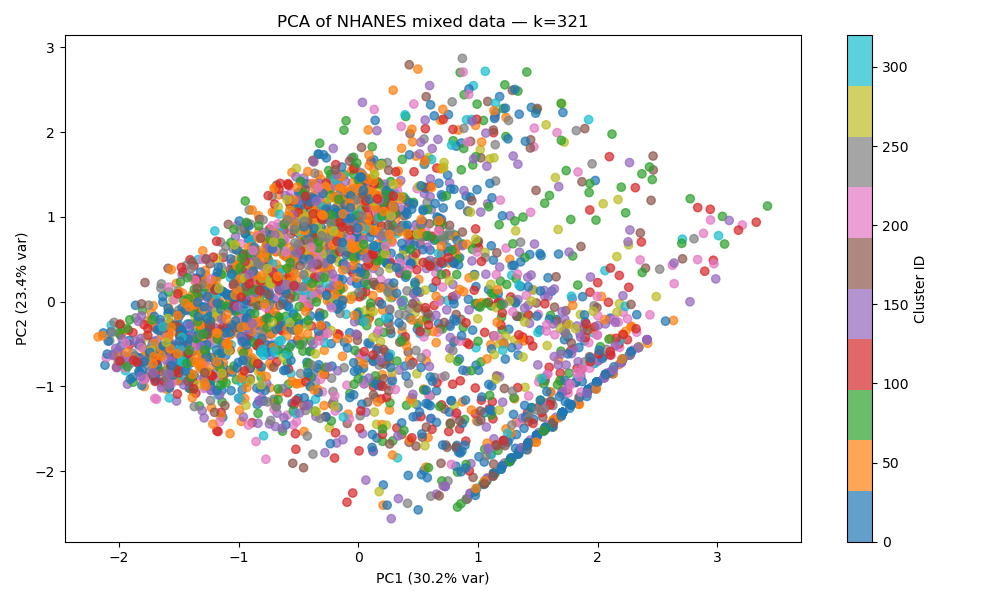}
    \caption{With Pregnancy Status}
    \label{fig:sub1}
  \end{subfigure}
  \quad
  \begin{subfigure}[b]{0.47\textwidth}
    \centering
    \includegraphics[width=\textwidth]{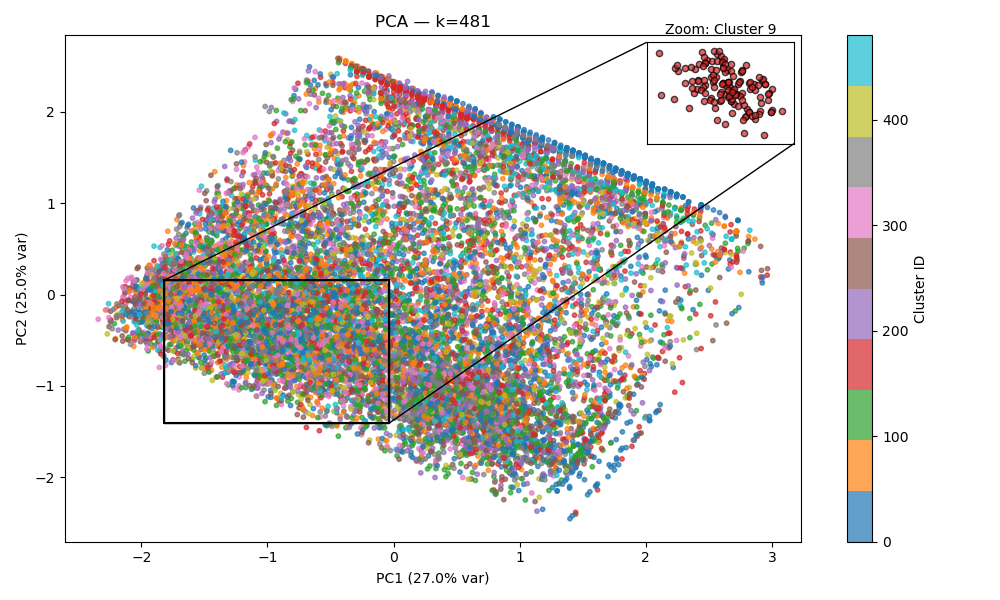}
    \caption{Without Pregnancy Status}
    \label{fig:sub2}
  \end{subfigure}

  \caption{PCA}
  \label{fig:PCA}
\end{figure}

We then did a PCA to see the contribution of each feature. Figure \ref{fig:PCA} shows the result.

\begin{figure}[htbp]
  \centering
  \begin{subfigure}[b]{0.47\textwidth}
    \centering
    \includegraphics[width=\textwidth]{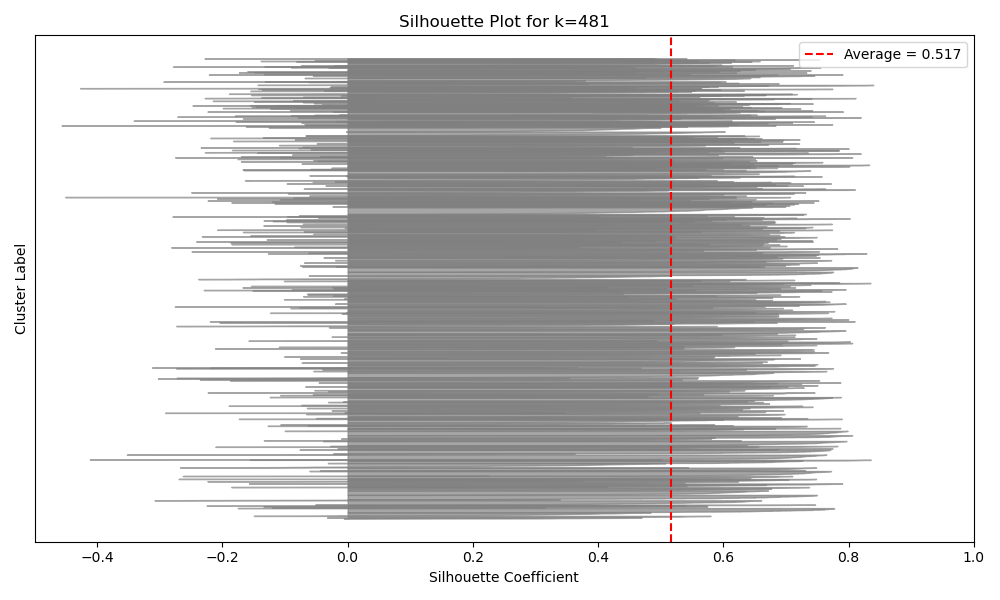}
    \caption{All Clusters}
    \label{fig:sub1}
  \end{subfigure}
  \quad
  \begin{subfigure}[b]{0.47\textwidth}
    \centering
    \includegraphics[width=\textwidth]{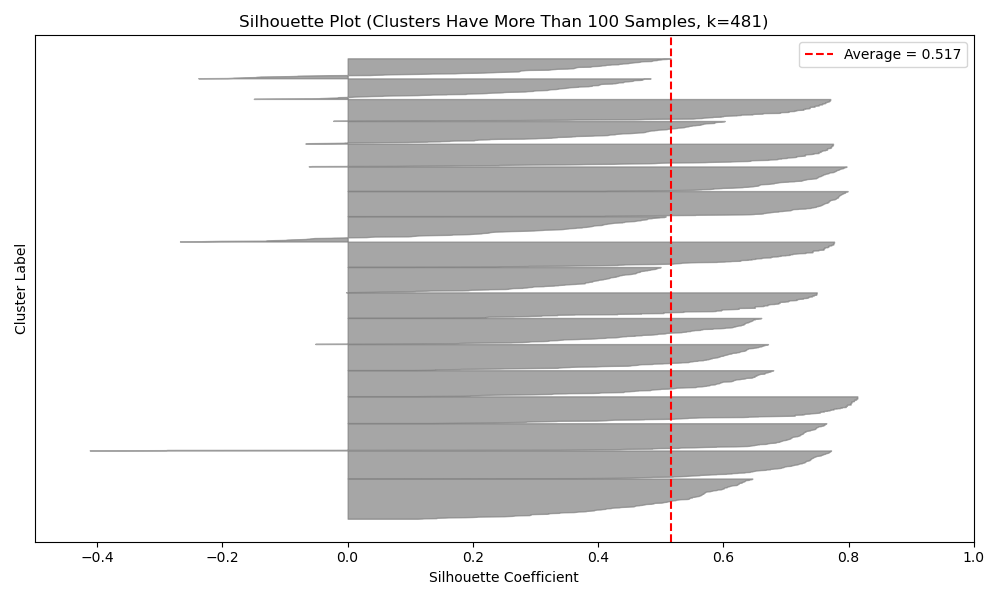}
    \caption{Clusters With $\geq$100 Samples}
    \label{fig:sub2}
  \end{subfigure}

  \vspace{1em}

  \begin{subfigure}[b]{0.31\textwidth}
    \centering
    \includegraphics[width=\textwidth]{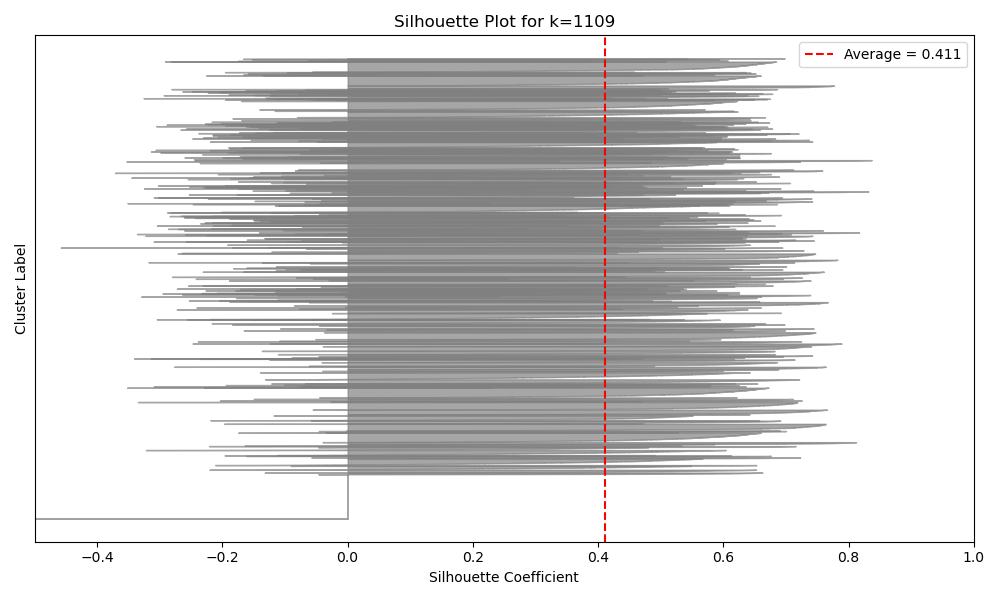}
    \caption{All Clusters\newline}
    \label{fig:sub3}
  \end{subfigure}
  \quad
  \begin{subfigure}[b]{0.31\textwidth}
    \centering
    \includegraphics[width=\textwidth]{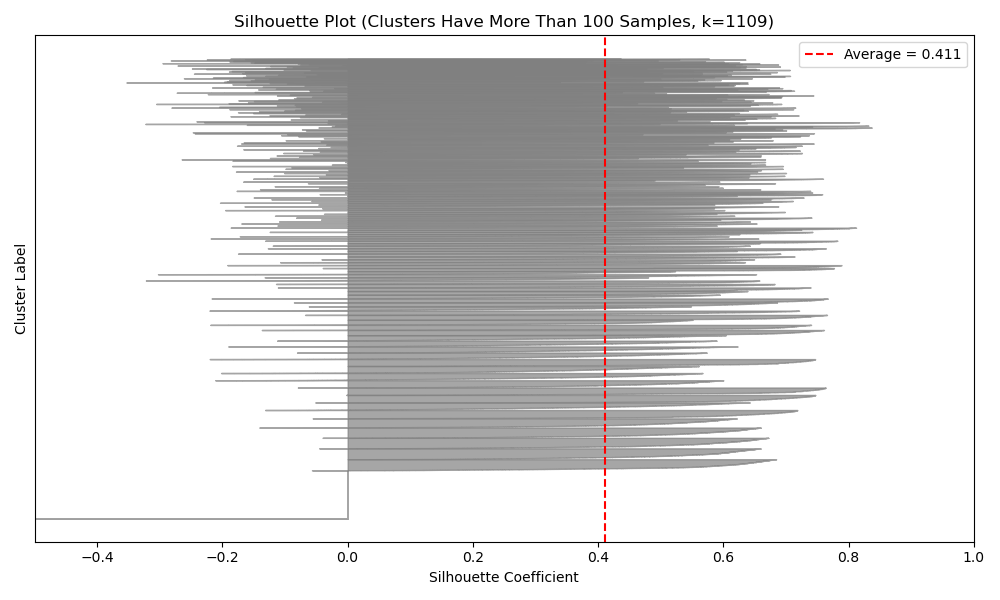}
    \caption{Clusters With $\geq$100 Samples}
    \label{fig:sub4}
  \end{subfigure}
  \quad
  \begin{subfigure}[b]{0.31\textwidth}
    \centering
    \includegraphics[width=\textwidth]{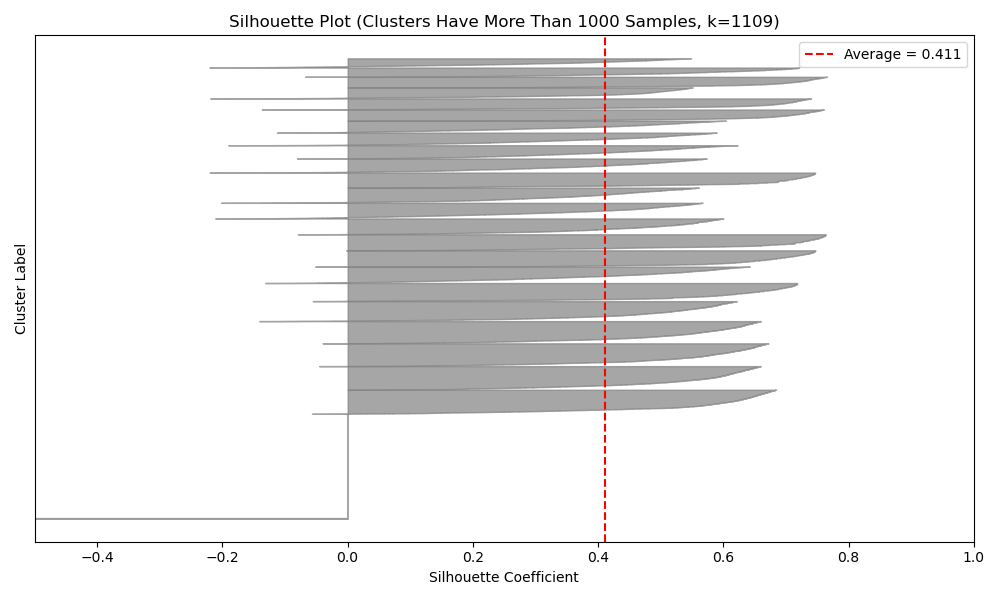}
    \caption{Clusters With $\geq$1,000 Samples}
    \label{fig:sub5}
  \end{subfigure}

  \caption{Average Silhouette Width With and Without the Pregnancy Status}
  \label{fig:avg_silhouette_coefficient}
\end{figure}

After that, we validated the resulting clusters via the average silhouette width. The average silhouette width (ASW) is used to measure the consistency of the clustering results. If the ASW is closer to 1, it indicates that the points of the cluster are all closer to its center and farther away from the other centers. If the ASW is negative, it means points are closer to a neighboring cluster than to their own. The results are shown as figure \ref{fig:avg_silhouette_coefficient}. It only includes records without pregnancy status. From figure \ref{fig:avg_silhouette_coefficient}, we can see most points have a positive ASW, and the total average ASW is 0.411, which indicates the model performance is good. After excluding invalid clusters, larger clusters exhibit a higher percentage of positive ASW, indicating that clusters with more samples achieve better separation.

\begin{figure}[htbp]
  \centering
  \begin{subfigure}[b]{0.31\textwidth}
    \centering
    \includegraphics[width=\textwidth]{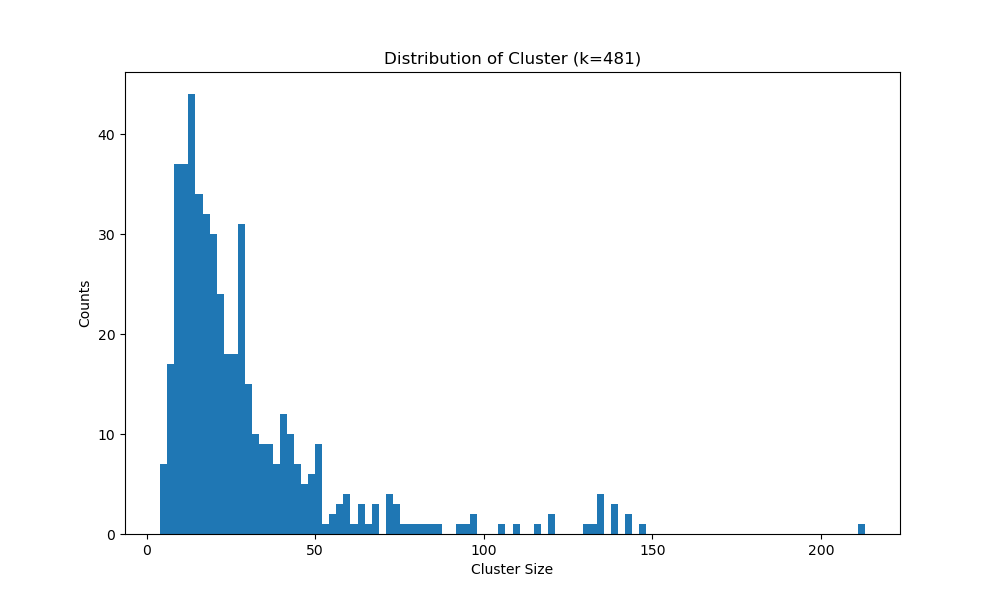}
    \caption{Remove NaN Values\newline}
    \label{fig:sub1}
  \end{subfigure}
  \quad
  \begin{subfigure}[b]{0.31\textwidth}
    \centering
    \includegraphics[width=\textwidth]{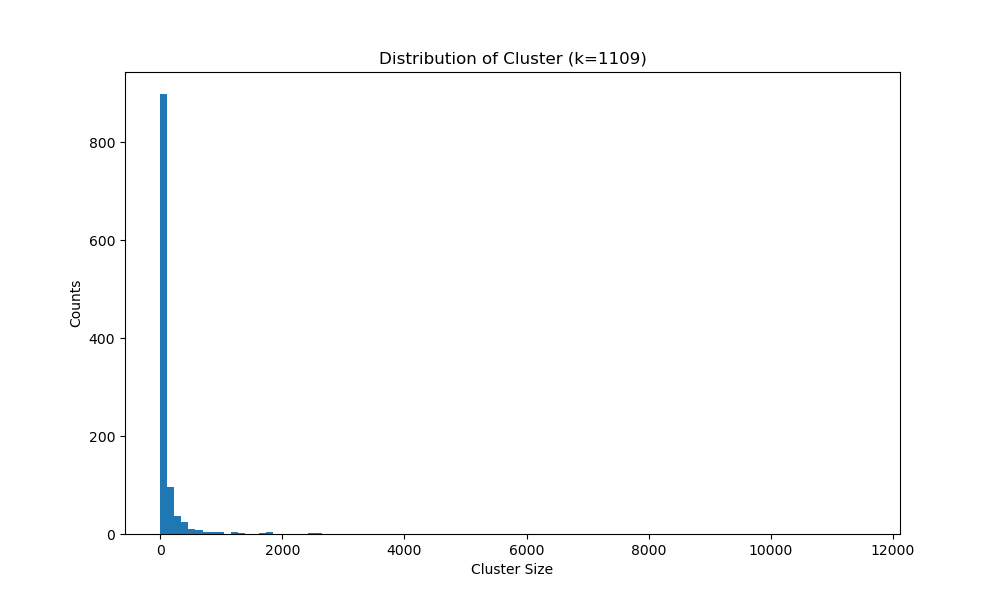}
    \caption{Retain NaN Values\newline}
    \label{fig:sub2}
  \end{subfigure}
  \quad
  \begin{subfigure}[b]{0.31\textwidth}
    \centering
    \includegraphics[width=\textwidth]{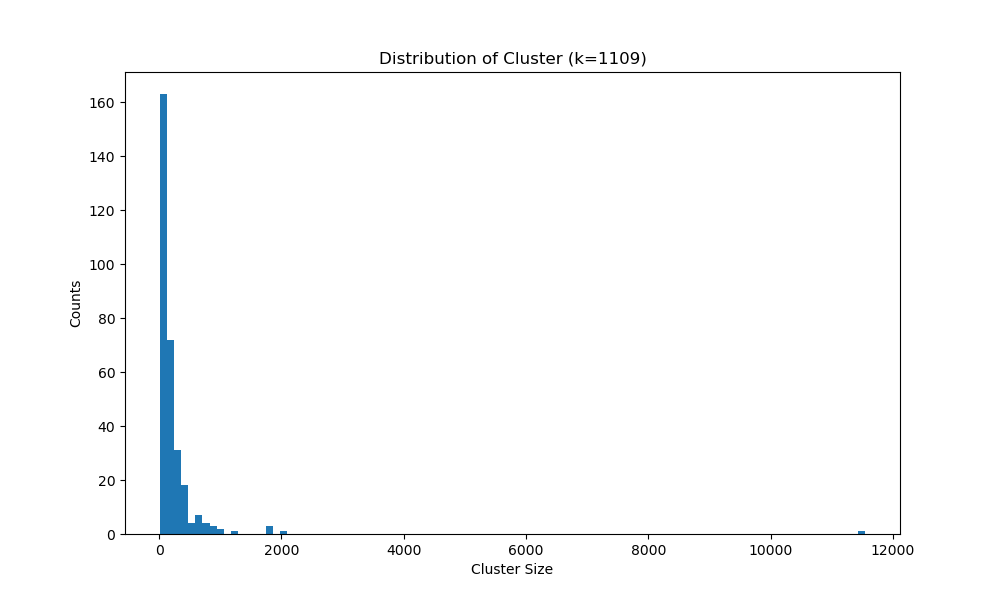}
    \caption{Retain NaN Values\newline Remove Invalid Clusters}
    \label{fig:sub3}
  \end{subfigure}

  \caption{Cluster Size Distribution}
  \label{fig:cluster_size_distribution}
\end{figure}

In the NaN-inclusive clustering, we dropped 740 single record clusters because at least one of their features was NaN and unable to form a valid status for the reinforcement learning model. We then removed 59 more clusters (41,881 records) to ensure every variable in the remaining clusters had at least one non-null value. After filtering, we got 310 valid clusters with 76,934 records. Figure \ref{fig:cluster_size_distribution} shows the distribution of cluster size.

\begin{figure}[htbp]
  \centering
  \begin{subfigure}[b]{0.47\textwidth}
    \centering
    \includegraphics[width=\textwidth]{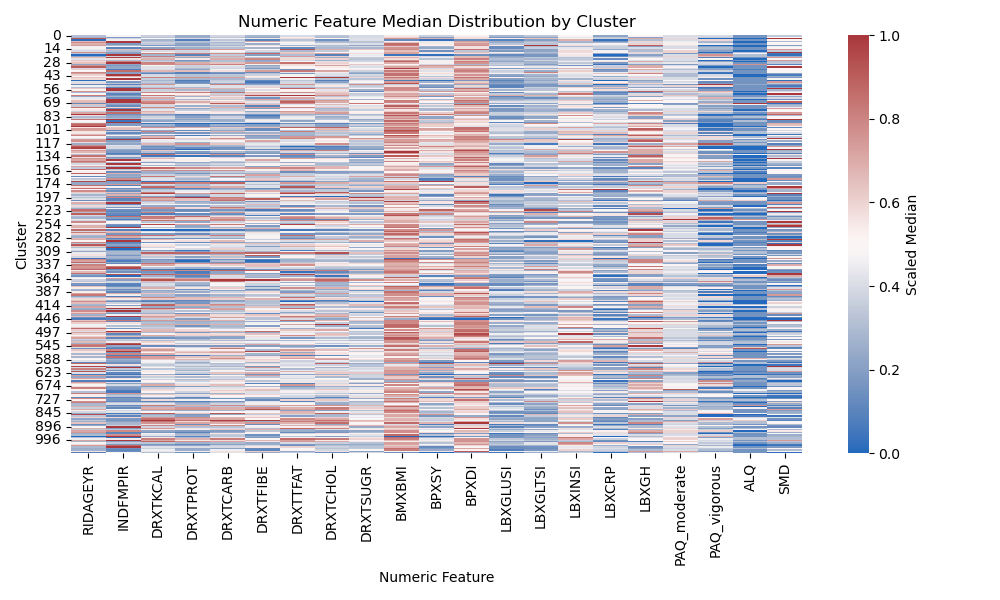}
    \caption{Numeric Feature Median Distribution}
    \label{fig:sub1}
  \end{subfigure}
  \quad
  \begin{subfigure}[b]{0.47\textwidth}
    \centering
    \includegraphics[width=\textwidth]{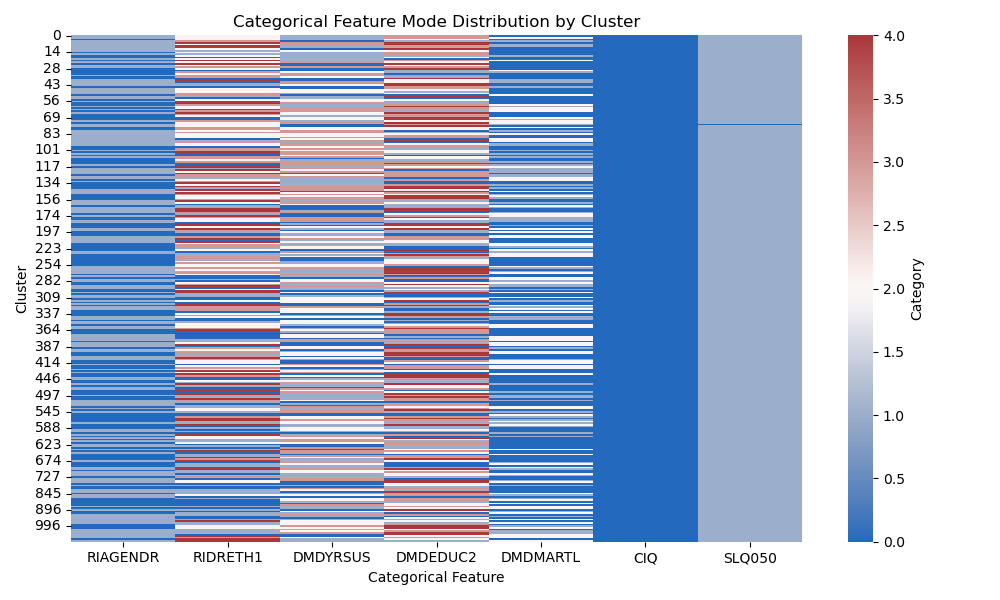}
    \caption{Categorical Feature Mode Distribution}
    \label{fig:sub2}
  \end{subfigure}
  \vspace{1em}
  \begin{subfigure}[b]{0.47\textwidth}
    \centering
    \includegraphics[width=\textwidth]{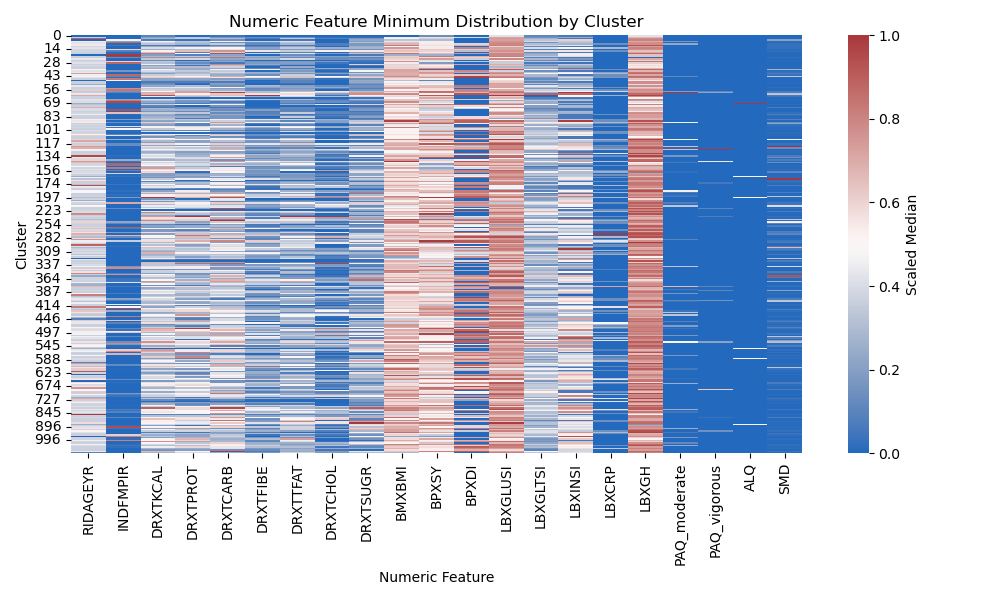}
    \caption{Numeric Feature Minimum Distribution}
    \label{fig:sub3}
  \end{subfigure}
  \quad
  \begin{subfigure}[b]{0.47\textwidth}
    \centering
    \includegraphics[width=\textwidth]{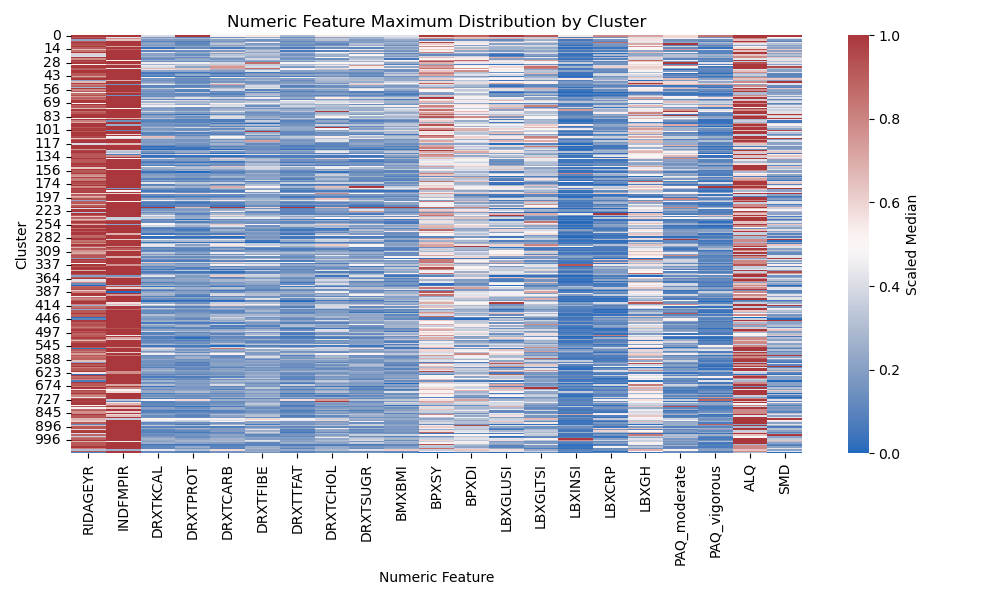}
    \caption{Numeric Feature Maximum Distribution}
    \label{fig:sub4}
  \end{subfigure}

  \caption{Feature Distribution by Cluster}
  \label{fig:feature_distribution_by_cluster}
\end{figure}

We also validated the feature distribution of each valid cluster. Figure \ref{fig:feature_distribution_by_cluster} shows the distribution of numeric and categorical features of all valid clusters. For numeric features, we reported the median, minimum, and maximum; for categorical features, we used the mode. As shown in Figure \ref{fig:feature_distribution_by_cluster}, numeric features vary distinctly across clusters, and categorical features perform the same except for depression and sleeping disorders.

\subsection{Reinforcement Learning Model}

Each cluster is then treated as one ``aggregated individual". Since participants can skip or refuse questions in NHANES interview, there is no record has all features complete in our dateset. To address this, we split each cluster into 4 quantile-based segments, treating each segment as a distinct status of an ``individual".

Let $\{x_1, x_2, \dot, x_i, \dot, x_N\}$ be our $N$ points, and let each $x_i$ be assigned to a cluster $\ell_i\in\{1,\dots,K\}$. Define
\[
  C_k = \bigl\{j : \ell_j = k\bigr\}
  \quad\text{for }k=1,\dots,K
\]
as the set of indices in cluster $k$. Let 
\[
  d^g(i,j)
\]
denotes the Gower distance between points $x_i$ and $x_j$. Then sum the Gower distance from point $x_i$ to all other points in its cluster:

\begin{equation}
\label{eq:cluster_gower_sum}
  s_i = \sum_{j \in C_{\ell_i}} d^g(i,j)
  \quad
  \bigl(i=1,\dots,N\bigr).
\end{equation}

Note that including or excluding the self‐distance $d(i,i)=0$ does not affect $s_i$. A smaller $s_i$ indicates that $x_i$ lies closer to its cluster members, making it a better representative of its cluster’s central tendency.

We then reordered the points within each cluster according to the Gower distance result and split them into 4 quantile-based segments. The empty values were filled in with median or mode within its cluster. Because the cluster sizes range from 19 records to 11,538 records, giving every cluster the same four values represents ``individual"'s status will over-represent small clusters and under-represent large ones relative to their true prevalence. Therefore, we allocated the number of values selected in each of the 4 quantile intervals proportional to cluster size. Define

\[
  w_i = \frac{S_i}{\sum_{k=1}^{K}S_k}
  \quad
  \bigl(i=1,\dots,K\bigr).
\]

where, $w_i\in (0, 1)$. $S_i$ refers to the size of cluster $\ell_i$. The allocate number of each cluster $c_i$ is calculated as:

\[
  c_i = max\bigl(1, 1240\times w_i\bigr)
  \quad
  \bigl(i=1,\dots,K\bigr).
\]

\begin{figure}[htbp]
  \centering
  \begin{subfigure}[b]{\textwidth}
    \centering
    \includegraphics[width=\textwidth]{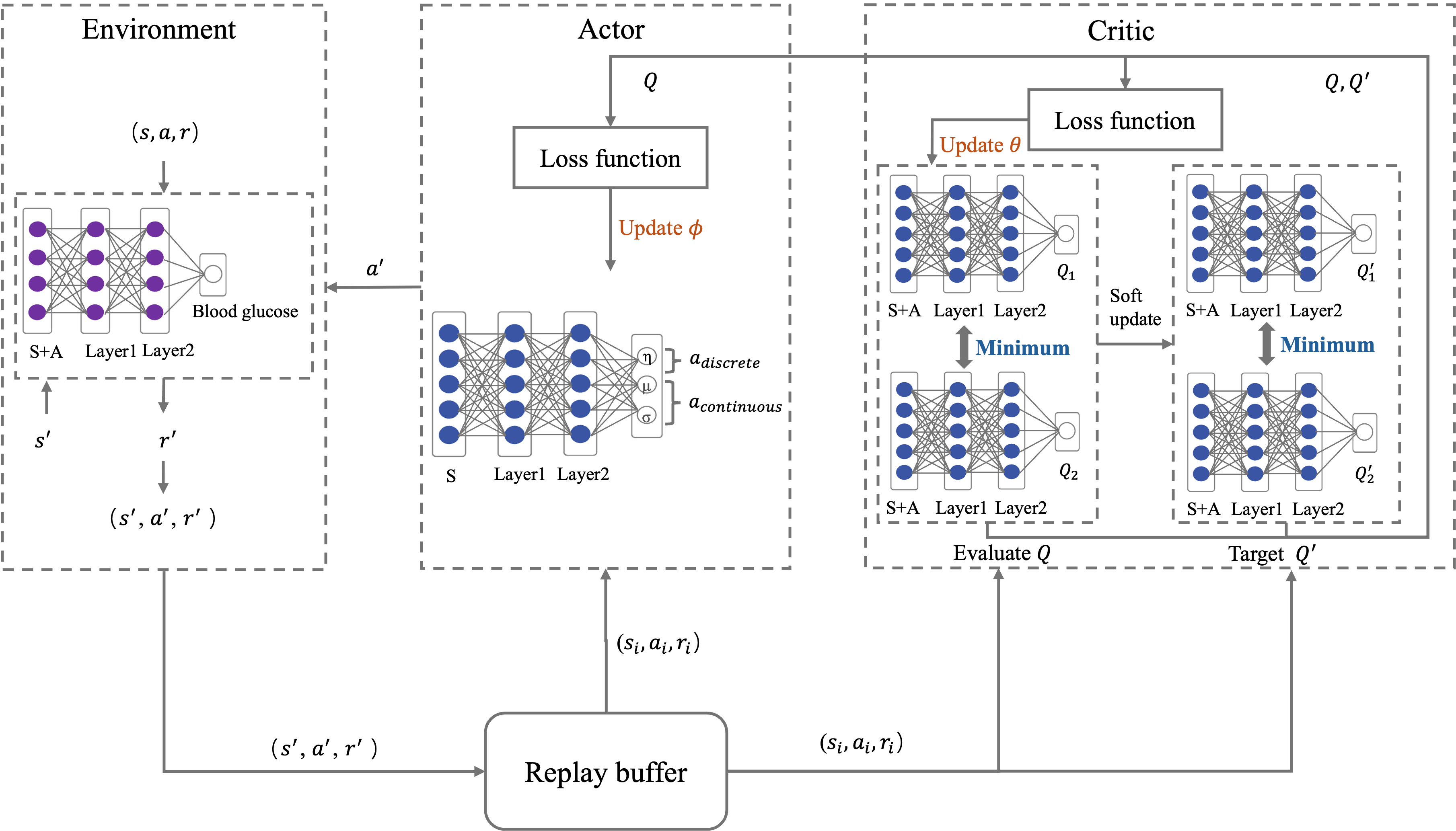}
  \end{subfigure}

  \caption{The Structure of Soft Actor Critic-Based Contextual Bandit Model}
  \label{fig:SAC_based_CB}
\end{figure}

Then, we had 3551 samples with action features, state features, and glucose value. Contextual bandit problem can be seen as a single-step Markov Decision Process (MDP) without temporal variables. Since the sample set doesn't have temporal components, we treated it as a contextual bandit problem. We see each NHANES scenario as a one-time snapshot of status and aims to choose the action that can immediately improve the glucose status. We used soft actor-critic algorithm (SAC) to solve this contextual bandit problem. Soft actor-critic is a maximum-entropy off-policy reinforcement learning algorithm, which the critic network learns to predict the immediate reward for state-actor pairs, and the actor network learns to pick one action that maximizes the trade-off between reward and entropy. Figure \ref{fig:SAC_based_CB} shows the structure of soft actor critic-based contextual bandit model.

\begin{figure}[htbp]
  \centering
  \begin{subfigure}[b]{0.47\textwidth}
    \centering
    \includegraphics[width=\textwidth]{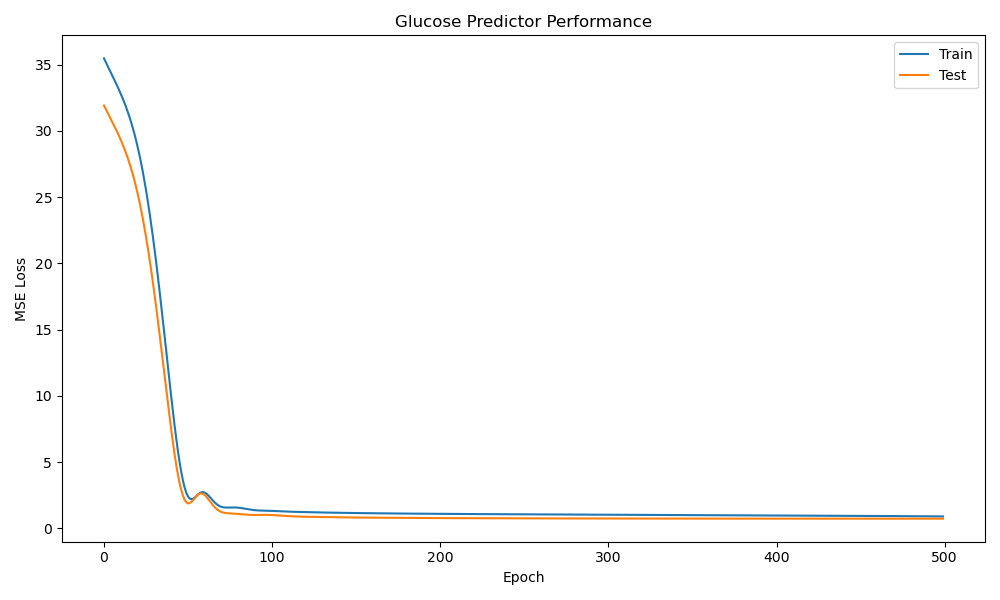}
    \caption{Environment Model MSE}
  \end{subfigure}
  \quad
  \begin{subfigure}[b]{0.47\textwidth}
    \centering
    \includegraphics[width=\textwidth]{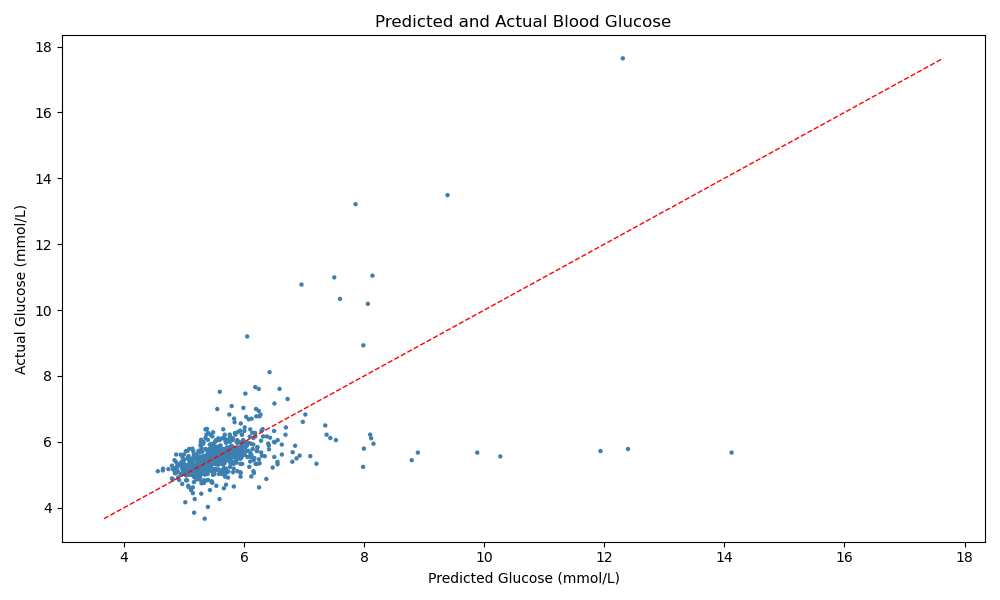}
    \caption{Predicted and Actual Blood Glucose}
    \label{fig:sub2}
  \end{subfigure}
  \caption{Environment Model Performance}
  \label{fig:environment_mse}
\end{figure}

For the environment model, we built a neural network with three fully connected layers to predict blood glucose under different combinations of states and actions. Figure \ref{fig:environment_mse} shows the mean squared error of the environment model under 1,000 iterations.

\begin{table}[ht]
\centering
\caption{Feature List}
\begin{tabular}{ccccc}
\toprule
\multicolumn{2}{c}{\textbf{Type}} & \textbf{State/Action} & \textbf{Feature} & \textbf{Explanation} \\
\midrule
\multicolumn{2}{c}{\multirow{18}{*}{Continuous}} & \multirow{7}{*}{State} & BMXBMI & Body Mass Index \\ 
& & & BPXSY & Blood pressure (systolic) \\
& & & BPXDI & Blood pressure (diastolic) \\
& & & LBXGLTSI & Two-hour glucose  \\
& & & LBXINSI & Insulin \\
& & & LBXCRP & C-reactive protein   \\
& & & LBXGH & Glycohemoglobin  \\
\cline{3-5}
& & \multirow{11}{*}{Action} & DRXTKCAL & Energy intake  \\
& & & DRXTPROT & Protein intake  \\
& & & DRXTCARB & Carbohydrate intake  \\
& & & DRXTFIBE & Fiber intake  \\
& & & DRXTTFAT & Fat intake  \\
& & & DRXTCHOL & Cholersterol intake  \\
& & & DRXTSUGR & Suger intake  \\
& & & ALQ & Drinking days in a year  \\
& & & $\text{PAQ}_{\text{moderate}}$ & Moderate activities  \\
& & & $\text{PAQ}_{\text{vigorous}}$ & Vigorous activities  \\
& & & SMD & \# cigarettes/month  \\
\midrule
\multirow{2}{*}{\multirow{9}{*}{Discrete}} & \multirow{3}{*}{Binary} & State & RIAGENDR & Gender \\
\cline{3-5}
& & \multirow{2}{*}{Action} & CIQ & Depression  \\
& & & SLQ050 & Sleeping disorder  \\
\cline{2-5}
& \multirow{6}{*}{Multi-Level} & \multirow{6}{*}{State} & RIDAGEYR & Age \\
& & & RIDRETH1 & Race  \\
& & & DMDYRSUS & Length of time stay in the U.S.  \\
& & & DMDEDUC2 & Education level \\
& & & DMDMARTL & Marital status  \\
& & & INDFMPIR & Family income  \\
\bottomrule
\end{tabular}
\label{tab:feature_list}
\end{table}

The features used are described in Table \ref{tab:feature_list}. Both state and action features have discrete and continuous variables. Therefore, we defined state $s$ as:

\[
s = (b_1, cat_1, cat_2,\dots, cat_m, cont_1, cont_2,\dots, cont_n) (m=6, n=7).
\]

where $b_1$ refers to the binary state parameter; $cat_1, cat_2,\dots,cat_m$ refers to 6 multi-level discrete state parameters; $cont_1, cont_2,\dots,cont_n$ refers to 7 continuous state parameters.

Similarly, we defined action $a$ as:

\[
a = (d_1, d_2, x_1, x_2,\dots,x_k) (k=11).
\]

where, $d_1 \in {0, 1}$ and $d_2 \in {0, 1}$ are the two binary action variables; $k$ is the number of continuous action parameters; $x_1, x_2,\dots, x_k \in \mathbb{R}$ is continuous action parameters. We assume each parameter in action is independent, and define the policy $\pi$ \cite{Xu2021} as:

\begin{equation}
\label{eq:policy}
\pi_\phi(a|s) = \pi_\phi(d_1|s)\,\pi_\phi(d_2|s)\,\pi_\phi(\mathbf{x}|s).
\end{equation}

For binary action parameter, the policy produces Bernoulli probabilities $\pi_\phi(d_1|s)$ and $\pi_\phi(d_2|s)$; for continuous action parameter, it follows a Gaussion distribution $\pi_\phi(\mathbf{x}|s)$.

In order to reduce bias and stabilize Q targets, we chose a double Q network for our SAC algorithm. Let $(s,a,r,s')$ be a transition tuple, $\pi_{\phi}(a'|s')$ be actor's next policy, $Q_{\bar\theta_1}$ and $Q_{\bar\theta_2}$ be target critics. Since contextual bandit problem only have one-step state, the tuple here becomes to $(s,a,r)$ and $\gamma$ is set to 0 because the contextual bandit problem focuses on immediate return. Then we have one-step target $y$:

\begin{equation}
\label{eq:target_y}
  y = r + \gamma[\text{min}(Q_{\bar\theta_1}(s',a'), Q_{\bar\theta_2}(s',a')) - \alpha\log\pi_\phi(a'|s')].
\end{equation}

where $r$ refers to the immediate return under the current state-action pair. Equation (\ref{eq:target_y}) therefore can be simplified to $y=r$. Therefore, we plug it in and have critic loss $\mathcal{L}{\text{critic}}$:

\[
\mathcal{L}{\text{critic}_i} = \mathbb{E}[\tfrac12\bigl(Q_{\theta_i}(s,a) - r\bigr)^2] (i\in \{1,2\}).
\]

The objective of actor is to maximize the expected reward and entropy, therefore, we want to maximize $V_{\pi_\phi}(s)$ under state $s$:

\begin{equation}
\label{eq:action_objective}
V_{\pi_\phi}(s) = \mathbb{E}[Q_{\theta}(s,a)] + \alpha\mathcal{H}(\pi_{\phi}(\cdot|s)).
\end{equation}

where the temperature parameter $\alpha$ is set to 0.2 in our experiment, which means we emphasize reward over exploration. The entropy term $\mathcal{H}$ is defined as:

\begin{equation}
\label{eq:entropy}
    \mathcal{H}(\pi_{\phi}(\cdot|s)) = -\mathbb{E}_{a\sim \pi_{\phi}}[\log \pi_{\phi}(a|s)].
\end{equation}

According to (\ref{eq:policy}), we have $\log \pi_{\phi}(a|s)$:

\begin{equation}
\label{eq:log_pi}
\log \pi_{\phi}(a|s) = \log \pi_{\phi}(d_1|s) + \log \pi_{\phi}(d_2|s) + \log \pi_{\phi}(\mathbf{x}|s).
\end{equation}

In (\ref{eq:log_pi}), it contains discrete and continuous action parameters $d_1, d_2, \mathbf{x}$. As mentioned before, Bernoulli and Normal distribution are produced accordingly. Therefore, we have (\ref{eq:dis_action}) refers to discrete action and (\ref{eq:cont_action}) refers to continuous action.

\begin{equation}
\label{eq:dis_action}
\pi_{\phi}(d_i|s) = \frac{e^{z_{i,j}(s)}}{\sum_{k}e^{z_{i,k}(s)}} (i\in \{0,1\}).
\end{equation}

where $z_{i,\cdot}(s)$ refers to logits for variable $i$.

\begin{equation}
\label{eq:cont_action}
\pi_{\phi}(\mathbf{x}|s) = \mathcal{N}(\mathbf{x}|\mu_\phi, \sigma_\phi^2).
\end{equation}

where, $\mu_\phi$ and $\sigma_\phi^2$ are the mean and variance value of sampling action $\mathbf{x}$ in state $s$.

Then we combined (\ref{eq:action_objective}), (\ref{eq:entropy}), (\ref{eq:log_pi}), (\ref{eq:dis_action}), and (\ref{eq:cont_action}) and finally got the actor loss $\mathcal{L}{\text{actor}}$:

\begin{equation}
\label{eq:actor_loss}
\mathcal{L}{\text{actor}} = \mathbb{E}[\alpha\log\pi_{\phi}(a|s) - min\{Q_{\theta_1}(s,a), Q_{\theta_2}(s,a)\}].
\end{equation}

\begin{figure}[htbp]
  \centering
  \begin{subfigure}[b]{\textwidth}
    \centering
    \includegraphics[width=\textwidth]{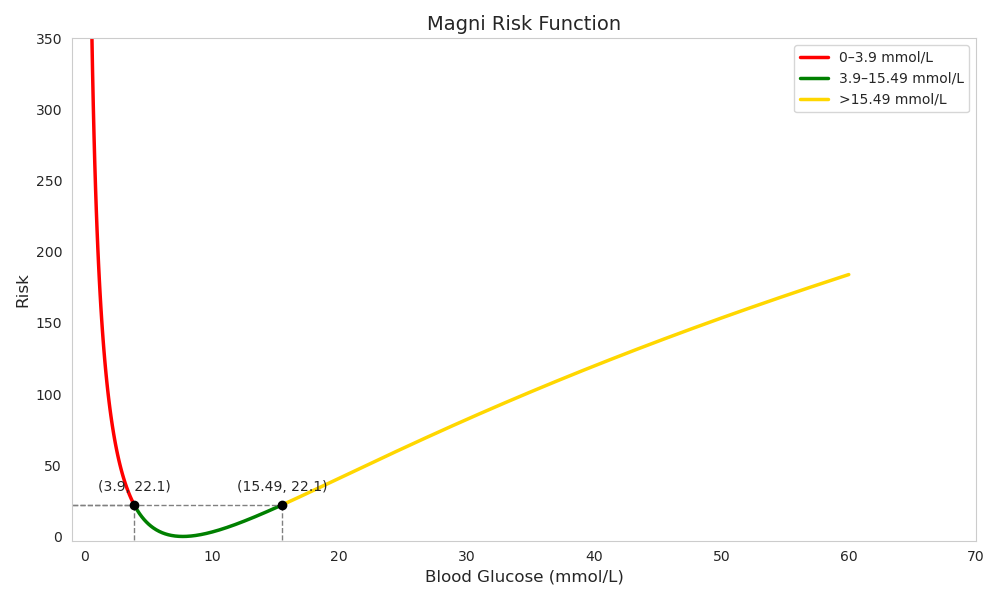}
  \end{subfigure}

  \caption{Magni Risk Function}
  \label{fig:magni_risk_function}
\end{figure}

Magni risk function \cite{Magni2007}, proposed by Magni in 2007, describes the risk to a person at different blood glucose values. It is defined as:

\[
  \text{risk}_i = 10 \times (c_0 \times (ln(BG)^{c_1} - c_2))^2.
\]

where, $c_0=3.35506$, $c_1=0.8353$, $c_2=3.7932$, $BG$ is the blood glucose (mmol/L). The Magni risk function with coefficients above penalizes hypoglycemia more severely than hyperglycemia. This is because hypoglycemia is more dangerous and life-threatening to people. Figure \ref{fig:magni_risk_function} shows the graphical representation of the Magni risk function. It presents the risk levels for different blood glucose.

We therefore defined the reward function as a negative Magni risk function:

\[
  \text{r}_i = -\text{risk}_i.
\]

\section{Result}

\begin{figure}[htbp]
  \centering
  \begin{subfigure}[b]{0.47\textwidth}
    \centering
    \includegraphics[width=\textwidth]{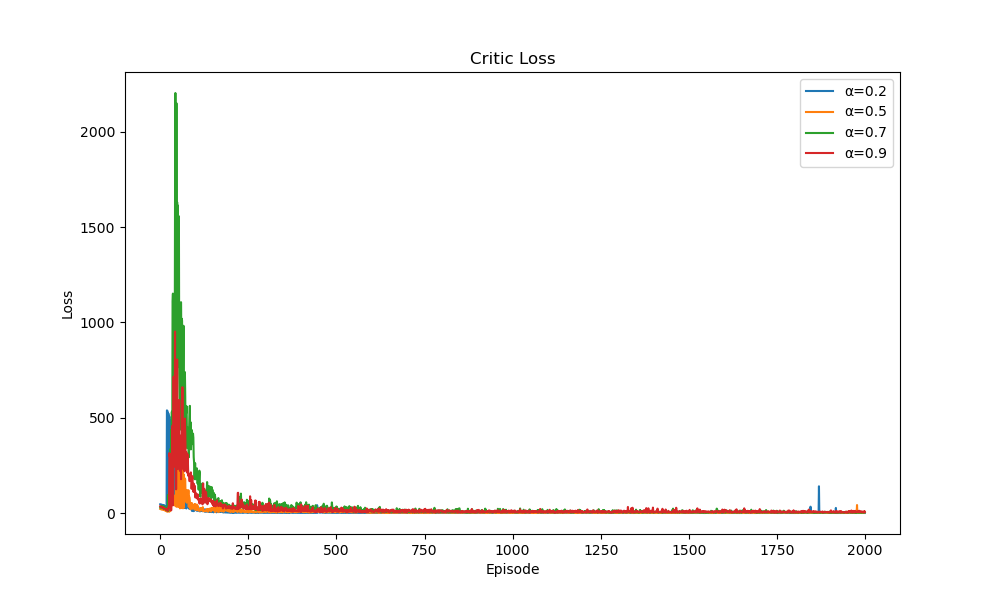}
  \end{subfigure}
  \quad
  \begin{subfigure}[b]{0.47\textwidth}
    \centering
    \includegraphics[width=\textwidth]{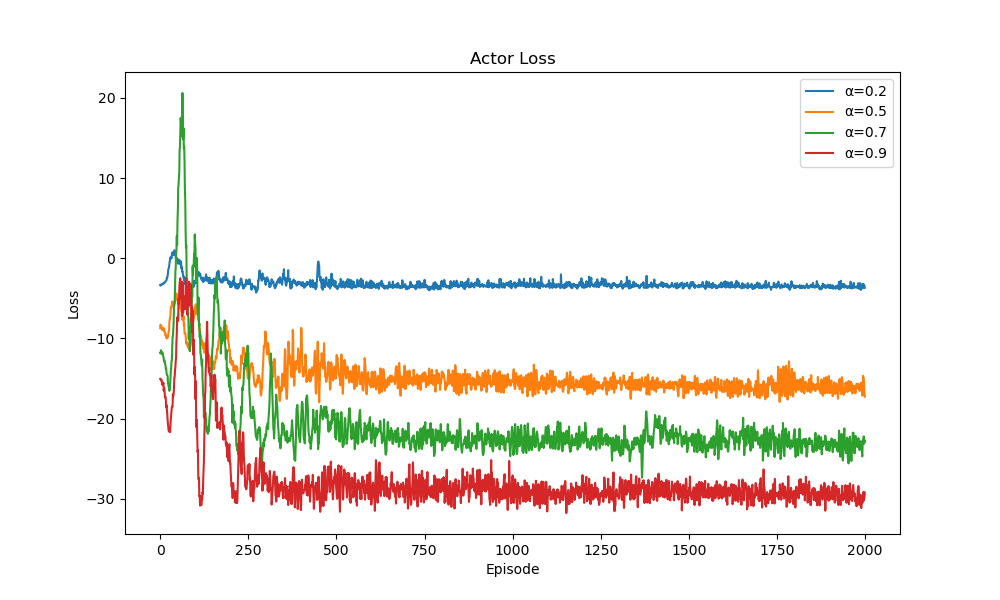}
  \end{subfigure}
  \caption{Actor and Critic Loss}
  \label{fig:sac_loss_comparsion}
\end{figure}

Figure \ref{fig:sac_loss_comparsion} shows the results of actor and critic loss under different $\alpha$. Since discrete action follows Bernoulli probability, which lies in the interval (0, 1), its natural log returns a non-positive value. As for continuous action, it follows Gaussian distribution:

\[p(x)=\frac{1}{\sqrt{2\pi}\,\sigma}\exp\Bigl(-\frac{(x-\mu)^2}{2\sigma^2}\Bigr).\]

Therefore, the log value of $p(x)$ is:

\[\log p(x) = -\frac{(x-\mu)^2}{2\sigma^2} - \ln\sigma - \tfrac12\ln(2\pi) = -\frac{(x-\mu)^2}{2\sigma^2} - \tfrac12\ln(2\pi\sigma^2).\]

Note that $-\frac{(x-\mu)^2}{2\sigma^2}$ is always non-positive, we focus on the term $-\tfrac12\ln(2\pi\sigma^2)$. We know that the term $2\pi\sigma^2 \geq 1$ when $\sigma\geq\frac{1}{\sqrt{2\pi}}$. Then, we have $-\tfrac12\ln(2\pi\sigma^2)\le0$ when $\sigma\geq\frac{1}{\sqrt{2\pi}}$. Therefore, $\log p(x) \leq 0$ when $\sigma\geq\frac{1}{\sqrt{2\pi}}$. Unless $\sigma$ has extremely small value or $x=\mu$, we know that $\log p(x)$ returns a non-positive value. As mentioned in equation \ref{eq:actor_loss}, the loss of actor therefore becomes more negative along with higher $\alpha$. It explains why in figure \ref{fig:sac_loss_comparsion}, the higher $\alpha$ shows a lower loss of actor.

\begin{figure}[htbp]
  \centering
  \begin{subfigure}[b]{0.47\textwidth}
    \centering
    \includegraphics[width=\textwidth]{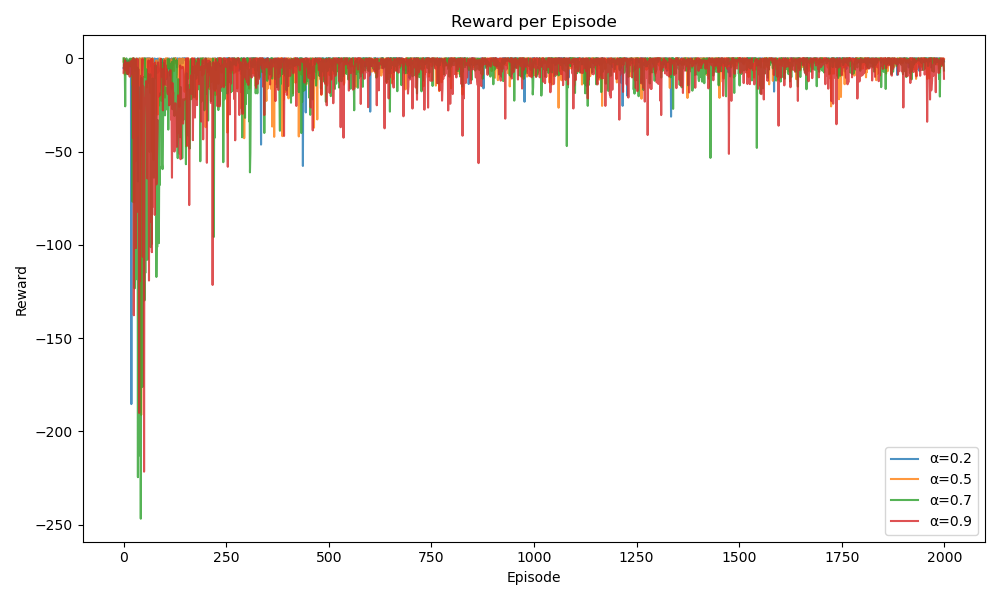}
    \caption{Reward}
    \label{fig:sub1}
  \end{subfigure}
  \quad
  \begin{subfigure}[b]{0.47\textwidth}
    \centering
    \includegraphics[width=\textwidth]{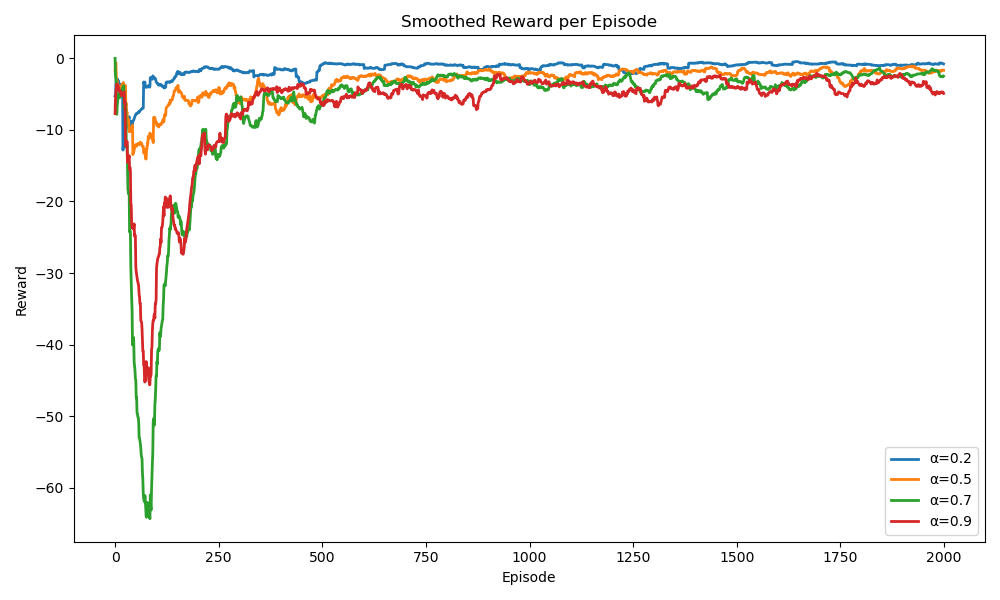}
    \caption{Smoothed Reward}
    \label{fig:sub2}
  \end{subfigure}

  \caption{SAC Reward with Different $\alpha$}
  \label{fig:SAC_reward_comparsion}
\end{figure}

Figure \ref{fig:SAC_reward_comparsion} shows the reward under different $\alpha$. We can see that although a high $\alpha$ yields a smaller actor loss, it is not consistent in reward performance. $\alpha$ controls the policy's exploration level and determines if put more probability mass on extreme actions. Since it is impossible to take a very extreme approach in the prescriptions in lifestyle medicine, a smaller value of a makes the most sense, both in practice and as a result of the figure.

\begin{figure}[htbp]
  \centering
  \begin{subfigure}[b]{0.47\textwidth}
    \centering
    \includegraphics[width=\textwidth]{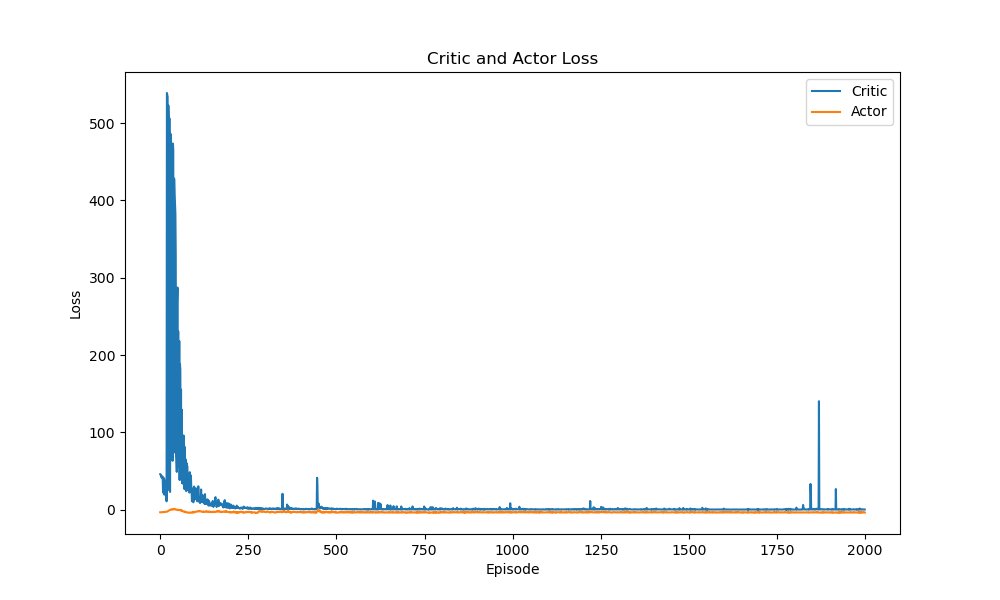}
    \caption{Loss}
  \end{subfigure}
  \quad
  \begin{subfigure}[b]{0.47\textwidth}
    \centering
    \includegraphics[width=\textwidth]{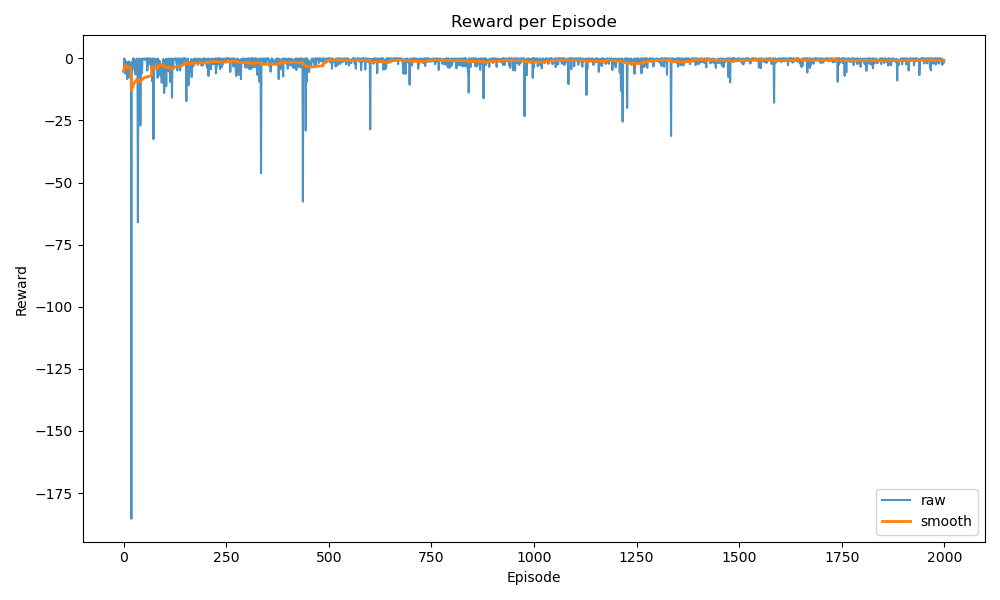}
    \caption{Reward}
  \end{subfigure}
  \caption{Model Performance When $\alpha=0.2$}
  \label{fig:sac_0.2}
\end{figure}

\begin{figure}[htbp]
  \centering
  \begin{subfigure}[b]{0.3\textwidth}
    \centering
    \includegraphics[width=\textwidth]{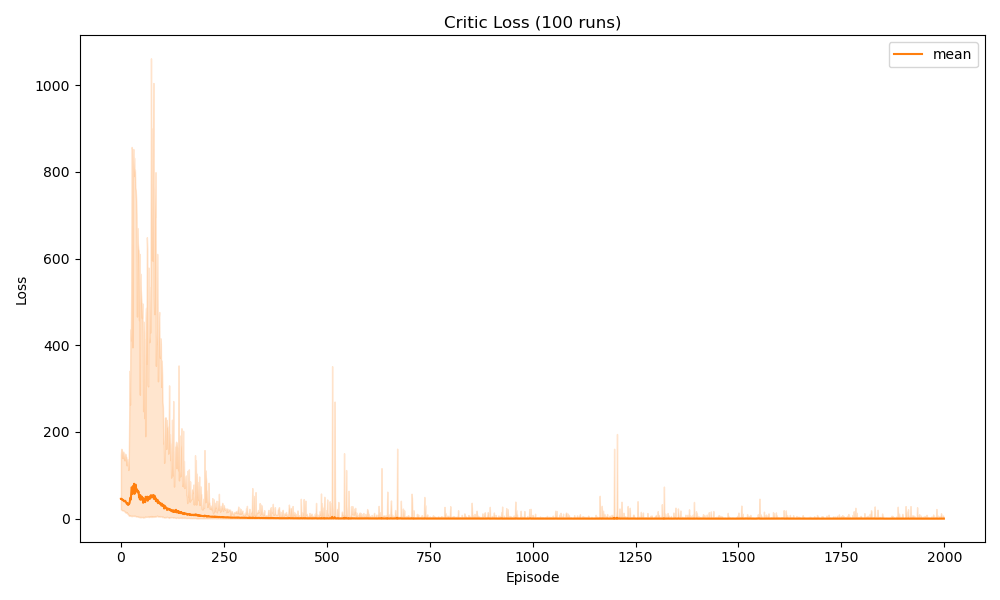}
    \caption{Critic Loss}
  \end{subfigure}
  \quad
  \begin{subfigure}[b]{0.3\textwidth}
    \centering
    \includegraphics[width=\textwidth]{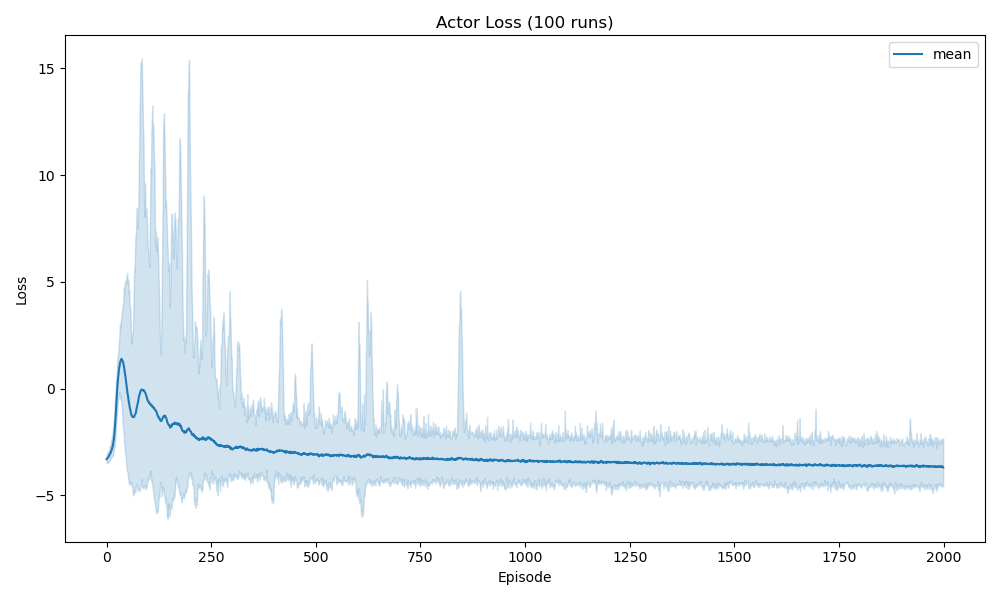}
    \caption{Actor Loss}
  \end{subfigure}
  \quad
  \begin{subfigure}[b]{0.3\textwidth}
    \centering
    \includegraphics[width=\textwidth]{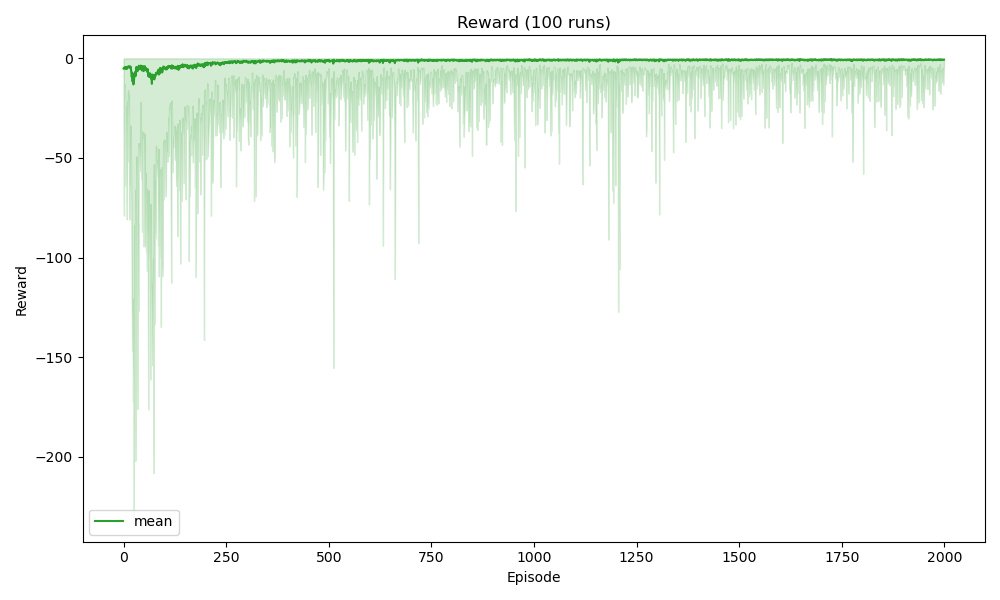}
    \caption{Reward}
  \end{subfigure}
  \caption{Model Performance When $\alpha=0.2$ in 100 Runs}
  \label{fig:sac_0.2_100runs}
\end{figure}

In our model, we chose $\alpha=0.2$. Figure \ref{fig:sac_0.2} shows the loss of actor and critic and reward under this setting. We can see both of them converge after 750 iterations. Since the reward function is the negative Magni risk function, all the reward values are negative. Higher reward means better performance. From figure \ref{fig:sac_0.2}, we can see that the reward value converge and does not exceed the dangerous range, which means the model satisfies our expectation. Figure \ref{fig:sac_0.2_100runs} shows the model performance under 100 runs. We can see that both losses and the reward are stable and converge to a low range after 500 iterations.

\section{Reference}
\bibliographystyle{abbrv}
\bibliography{LR}

\end{document}